\documentclass{tJOT2e}
\newcommand{\ov}{\overline}
\newcommand{\tl}{\tilde}
\newcommand{\bs}{\boldsymbol}
\usepackage{graphics,psfrag,epsfig,rotating}
\usepackage[monochrome]{color}
\usepackage{subfigure}
\usepackage{amsmath,amssymb,amstext}
\nopagebreak 
\begin{document}
\title{A derivation of the NS-$\alpha$ model and preliminary application to plane channel flow}
\author{K. Andrea Scott$^{\dagger}$$^{\ast}$\thanks{$^\ast$Corresponding author. Email: ka3scott@engmail.uwaterloo.ca} \vspace{6pt} and F.S. Lien $^{\dagger}$\\\vspace{6pt}  $^{\dagger}${Department of Mechanical Engineering, University of Waterloo, 200 University Avenue West, Waterloo, Canada N2L 3G1}};
\maketitle
\section{Abstract}
In this paper the Navier-Stokes-$\alpha$ (NS-$\alpha$) model is considered within a large-eddy simulation framework. An investigation is carried out  using fully-developed turbulent channel flow at a fairly low Reynolds number. This is a flow where diffusion plays a prominent role, and presents a challenge to the nonlinear model investigated here. It is found that when $\alpha^{2}_{k}$ is based on the mesh spacing, the NS-$\alpha$ model has a tendency to tilt spanwise vorticity in the streamwise direction, leading to high skin friction. This is due to interaction between the spanwise vorticity, the model, and the streamwise streaks. To overcome this problem $\alpha^{2}_{k}$ is damped in the streak affected region. Results overall demonstrate the potential of the model to reproduce some features of the DNS (helicity statistics and small-scale features), but more work is required before the full potential of the model can be achieved. In addition to the channel flow investigation, a derivation of the governing equations using Hamilton's principle is given. The derivation is intended to be clear and accessible to a wide audience, and contains a new interpretation of the model parameter. 

\begin{keywords} Navier-Stokes-alpha; large-eddy simulation; plane channel flow; regularization model; subgrid-scale model

\end{keywords}\bigskip

\section{Introduction}

Traditionally, turbulence models are derived by applying averaging (RANS) or filtering (LES) techniques to the Navier-Stokes equations. This results in a momentum equation with an unclosed term, known as the Reynolds stress or subgrid stress, that must be modeled. Numerous models have been proposed over the years \cite{Geurts2003, Piomelli1999,Hanjalic1994}. The majority of these models employ an eddy viscosity ansatz. This is well-founded in the sense that an eddy viscosity is a reasonable model for the energy drain provided by the small scales that have been removed during the filtering or averaging procedure, and is popular in part because adding viscosity generally renders a simulation more stable. However, the shortcomings of the eddy viscosity approach are well known. Linear eddy viscosity models employ a simple constitutive relationship that assumes alignment between the subgrid stress, $\tau_{ij}$,  and the strain rate. It has been found both in analysis of DNS data and in experimental studies that this is far from the truth \cite{Tao2002}.  Eddy viscosity models are also strictly dissipative for positive viscosities, and unstable for negative ones.  This means they cannot capture the reverse energy transfer from small to large scales, known as backscatter. Although energy transfer is on average from large to small scales in three-dimensional turbulence, there are a number of flows where local backscatter effects are important. Examples from shear flows include the later stages of boundary layer transition \cite{Piomelli1991}, hairpin vortices in the near-wall region \cite{Piomelli1996,Hartel1994} and vortex pairing in mixing layers \cite{Silva2002}. A popular method to incorporate backscatter is by adding a stochastic forcing term to the eddy viscosity model \cite{Mason1994}. The rationale behind this is that while a dissipative model can capture the mean forward transfer, the subgrid stress exhibits significant fluctuations about this mean, and it is these fluctuations that are responsible for the backscatter. In practice though, backscatter tends to be strongly correlated with coherent structures, leading some to hypothesize that a deterministic model may be more appropriate \cite{Piomelli1996}.  
\\
\\
One flow where traditional eddy viscosity models have difficulty is turbulent channel flow. In this case, the eddy viscosity needs to be reduced close to the wall, or a dynamic procedure needs to be used, to avoid damping out the turbulence. In this paper we study the turbulent channel flow using the NS-$\alpha$ model. The NS-$\alpha$ model is different from an eddy viscosity model in that instead of adding an eddy viscosity term to a filtered momentum equation, it is a nonlinear regularization. The model can be thought of most intuitively as a vorticity regularization. The vorticity equation for the inviscid form of the NS-$\alpha$ model is (in this paper the use of repeated indices implies a summation, unless otherwise stated),
\begin{equation}
\frac{\partial \omega_{i}}{\partial t}+\tl{u}_{j}\frac{\partial \omega_{i}}{\partial x_{j}}=\omega_{j}\frac{\partial \tl{u}_{i}}{\partial x_{j}}.
\label{NSalphavort}
\end{equation}
In this equation the background flow is smoothed, which means the velocity gradients  become less effective at stretching and tilting the vortices. In turn this suppresses the forward transfer of energy and prevents the creation of smaller and smaller scales, hence eliminating the need to model the effects of these scales when we carry out a coarse grid numerical simulation. In this way additional viscosity is not needed, per se. This phenomenological view of the model is supported by Fourier transform analysis of the nonlinear terms \cite{Domaradzki2001,Scott2008}, which shows that if the underlying dynamics follow those of the Navier-Stokes equations, then  the system stops transferring energy \textit{to} small scales when a certain wavenumber, say $k_{\alpha}$, is exceeded. At the same time, it does not stop the backscatter. Thus we do not need to model the \textit{missing} backscatter because it has not been removed.  
\\
\\
In spite of its intuitive application as a turbulence model \cite{Holm1999,Chen1998,Chen1999a,Chen1999b,Foias2001} , there have been only a few attempts in the literature to use the NS-$\alpha$ model outside of idealized box turbulence experiments \cite{Graham2007,Graham2008,Chen1999b,Mohseni2003}. Geurts and Holm \cite{Geurts2006} used the model to capture temporal transition in a mixing layer, while Holm and Nadiga \cite{Holm2003} were able to produce a four-gyre structure on a coarse mesh that would only produce two gyres when viscosity was used as the closure.  Recently, the NS-$\alpha$ model was incorporated into a primitive-equation ocean model  where it was found to produce energetic eddies at a coarse resolution where eddy viscosity approaches failed \cite{Petersen2008}. In all of these studies they maintained a constant value for the model parameter $\alpha$, as a fraction of the mesh spacing. This was done because $\alpha$ relates the smoothed and unsmoothed velocities and can be interpreted as a filter width. However, we expect in flows which are highly anisotropic, such as near wall flows, that we will not be able to maintain a constant value of $\alpha$. This topic was explored by Scott and Lien \cite{Scott2010} the NS-$\alpha$ model was applied to a lid-driven cavity flow, and both $\alpha$ as a function of the mesh spacing and a flow dependent version were tested. The flow dependent version was found to reproduce the results of the DNS fairly well. 
\\
\\
Another study that used a non-constant $\alpha$ was an investigation of turbulent channel flow  by Zhao and Mohseni \cite{Zhao2005a}, where a dynamic version of the NS-$\alpha$ model was developed and tested in an a priori manner, in which $\alpha^{2}$ was calculated, but was not fed back to the flow. A later investigation of the channel flow where the model was tested \textit{a posteriori} found the model to produce high spanwise fluctuations \cite{Zhao2005b}.  One of the objectives of the present study was to determine the source of this problem. We will show here in section \ref{modelbias} that this occurs because the model has a bias towards tilting vorticity in the streamwise direction close to solid walls. Our response to this bias is given in section \ref{biasresp}. 
\\
\\
Before this, a derivation of the governing equations is presented in Section \ref{derivation}, followed by the formulation of the subgrid model and description of the test case. The derivation is comprised in part of principles seen before \cite{Bretherton1970,Holm1999,Bhat2005}, and is intended to be more accessible to the non-mathematician than that commonly encountered in the literature \cite{Holm1999,Chen1999c,Marsden2003,Soward2008}.  Unlike previously published derivations, the current one shows explicitly the steps used in varying the action, and we hope it will serve as a useful basis for further extensions of the model to different flows. The derivation also contains a new interpretation of the model parameter, showing how it can be related to a particle displacement error, and shows explicitly how the variation with respect to this parameter is carried out. 
%A later study where the model was test a posteriori for the channel flow did not prove to be as good. Their dynamic model was based on the assumption that the subgrid stress $\tau_{ij}=\ov{u_{i}u_{j}}-\ov{u}_{i}\ov{u}_{j}$ can be \textit{modelled} using the LANS subgrid stress, $m_{ij}$. However, there is some inconsistency in this approach because in fact the LANS subgrid stress is already an exact model for $insert term here$. 

\section{Derivation}
\label{derivation}
\noindent
The NS-$\alpha$ equations differ from other approaches to turbulence modelling in that  the effects of turbulence are introduced at the level of the variational principle. This is done here using Hamilton's principle, which is a variational principle that leads to Newton's second law. To incorporate the effects of turbulence within the framework of Hamilton's principle, consider that in the material description of a fluid, the state of a fluid particle with label $\bs{a}$ is specified by the particle displacement $\boldsymbol{x}(\boldsymbol{a},t)$, and velocity, $\dot{\boldsymbol{x}}(\bs{a},t)$. A momentum equation arises when the first variation of the action
\begin{equation}
S[\boldsymbol{x},\dot{\boldsymbol{x}}]=\int_{t_{1}}^{t_{2}}\int_{V(\bs{a})}l_{m}(\boldsymbol{x},\dot{\boldsymbol{x}})d^{3}a \, dt,
\label{material}
\end{equation} 
is set to zero. In \eqref{material} $l_{m}$ is the material representation of the Lagrangian density (Lagrangian/unit volume), and is the difference between the kinetic energy $T$ and the potential energy $\Phi$.  Turbulence can be incorporated within this framework by adding a random component to the displacement of a fluid particle, and, given that this random component is a function of time, to its velocity. This is the method pursued by Marsden and Shkoller \cite{Marsden2003} and Holm \cite{Holm1999}.
\\
\\
In the following we will pursue a different approach and work with the Eulerian description of a fluid, where the state of the (isentropic) fluid is described by $\boldsymbol{u}(\boldsymbol{x},t)$ and $\rho(\boldsymbol{x},t)$. This is the description that is usually used in developing a turbulence model. For example, the well-known RANS equations for an incompressible flow are developed by decomposing the velocity field into mean and fluctuating components, substituting into the Navier-Stokes equations, and averaging. To facilitate the derivation of the model we have split it into the five sections. Here a brief description of each section is given.
\begin{itemize}
\item \textit{Definition of the Lagrangian} In this section we describe the particular form of the Lagrangian we are using, which is for constant density, incompressible flow with no sources of potential energy. To apply the model to different flows, the definition of the Lagrangian must be modified. Here we write the Lagrangian in Eulerian coordinates, but it is also possible to work with material coordinates.
\item \textit {Incorporation of turbulence into the Lagrangian} In this section we describe the definition of the velocity fluctuation. This is the only approximation that enters the NS-$\alpha$ model. This section follows that given in Holm \cite{Holm1999}, with a different interpretation of the model parameter.
\item \textit{Varying the action} In this section the first variation of the action is taken. This is an application of the calculus of variations. If different boundary conditions are used for the model parameter, this part should be modified.
\item \textit{Definition of the variations of the Eulerian coordinates} In this section the relationship between a trajectory variation and the Eulerian variables are used to define their variations. This section is particular to the use of Eulerian coordinates.
\item \textit{Setting the variation to zero}.  When the first variation is set to zero, we arrive at the momentum equation. 
\end{itemize}

\subsection{Definition of the Lagrangian}
In Eulerian coordinates the action principle is (c.f. \cite{Bretherton1970,Salmon1998}),
\begin{equation}
S[\boldsymbol{u},\rho,s]=\int_{t_{1}}^{t_{2}} \int_{V(\bs{x})}l(\boldsymbol{u},\rho,s)d^{3}x \, dt.
\label{spatial}
\end{equation}
with Lagrangian density 
\begin{equation}
l=
\frac{\rho(\bs{x},t)}{2}u_{i}(\bs{x},t)u_{i}(\bs{x},t)
-E(\rho(\bs{x},t),s(\bs{x},t))-V(\bs{x},t)
\end{equation}
where $E$ is the internal energy, $s$ is the entropy and $\phi$ is the potential energy. Here we will consider an incompressible, constant density fluid of uniform entropy with no sources of potential energy. In the action principle we then remove the internal and potential energy functions and add an equation constraining the density to be constant
\begin{equation}
S_{\rho}=\int_{t_{1}}^{t_{2}}\int_{V}\left(
\frac{\rho(\bs{x},t)}{2}u_{i}(\bs{x},t)u_{i}(\bs{x},t)
+p\left(\rho_{o}-\rho(\bs{x},t)\right)
\right)d^{3}x\,dt.
\end{equation}
Here $p$ is the pressure, and is a Lagrangian multiplier.  Using conservation of mass we can relate the density ratio to the volume element $D$ by $D=\rho/\rho_{o}$, where $\rho_{o}$ is the reference density. The volume element is defined as \cite{Holm1999} the ratio of the volume in the initial configuration to that in the current configuration
\begin{equation}
D=det (d \bs{a}/d \bs{x})
\end{equation}
Using this we arrive at the action 
\begin{equation}
S=\int_{t_{1}}^{t_{2}}\int_{V}\left(
\frac{D}{2} u_{i}(\boldsymbol{x},t)u_{i}(\boldsymbol{x},t)+p\left(1-D\right)\right)d^{3}x \, dt,
\label{action}
\end{equation}
where we have divided through by $\rho_{o}$.

\subsection{Incorporation of turbulence into the Lagrangian}
To incorporate turbulence the velocity is then expressed as the sum of a mean component and a random fluctuation, in a similar manner as what is done in RANS (here $\phi$ is a random variable, or a fast time-scale \cite{Holm1999})
\begin{equation}
u_{i}(\boldsymbol{x},t;\phi)=\ov{u}_{i}(\boldsymbol{x},t)+{u}_{i}'(\boldsymbol{x},t;\phi),
\label{vdecomp}
\end{equation}
where the averaging operator $\langle \cdot \rangle$ and $\ov{\bs{u}}$ are defined as \cite{Holm1999}
\begin{equation}
\ov{\bs{u}}(\bs{x},t)=\langle \boldsymbol{u} (\bs{x},t;\omega) \rangle=\lim_{T \to \infty} \frac{1}{T} \int_{0}^{T} 
\boldsymbol{u}(\boldsymbol{x},t;\omega) \, d\omega.
%or show this as a summation over n realizations of a random variable
\end{equation}
The only approximation in the NS-$\alpha$ model comes in the definition of the velocity fluctuation. For example, we can write (to first order)
\begin{equation}
\ov{u}_{i}(\bs{x}+\bs{\xi})=\ov{u}_{i}(\bs{x})+\xi_{j}\frac{\partial \ov{u}_{i}}{\partial x_{j}}.
\label{tsvfluc}
\end{equation}
Defining the velocity fluctuation as the difference between our averaged velocity at two points, $\bs{\ov{u}}(\bs{x})$ and $\bs{\ov{u}}(\bs{x}+\bs{\xi})$, gives
\begin{equation}
u'_{i}=-\xi_{j}\frac{\partial \ov{u}_{i}}{\partial x_{j}}.
\label{vfluc}
\end{equation}
For example, we can consider that if we are sitting at a field point $\bs{x}$ occupied by a particle with velocity $\bs{u}(\bs{x})$ at time $t$, and then at a later time our field point is occupied by a particle that was previously at $\bs{x}+\bs{\xi}$, and has velocity $\bs{u}(\bs{x}+\bs{\xi})$, the velocity fluctuation is then given by \eqref{vfluc}. The same expression for the velocity fluctuation can be derived in a similar manner by expanding Eulerian and Lagrangian velocities in terms of $\xi$ \cite{Holm1999}.
%For example, we can consider that if we are sitting at a field point $\bs{x}$ occupied by a particle with velocity $\bs{u}(\bs{x})$, but there is some uncertainly related to the particle velocity that is reflected in our resolution, for example whether the particle is at a position $\bs{x}$ or a position $\bs{x}+\bs{\xi}$. This uncertainty is reflected in the expression for the velocity fluctuation, given by \eqref{vfluc}. The same expression for the velocity fluctuation can be derived in a similar manner by expanding Eulerian and Lagrangian velocities in terms of $\xi$ \cite{Holm1999}.
\\
\\
An alternative interpretation of this picture can be found by looking at $\bs{\xi}$ as the error between true and modelled trajectories. Note that in this discussion on the interpretation of the model parameter as an error there is no summation on repeated indices. Given that
\begin{equation}
\frac{D\boldsymbol{x}}{Dt}=u,
\end{equation}
a simple first order discretization of the true trajectory $\boldsymbol{x}_{t}$ and the modelled trajectory $\boldsymbol{x}_{m}$ would be
\begin{equation}\begin{split}
\boldsymbol{x}^{n}_{t}=\boldsymbol{x}^{n-1}_{t}+\boldsymbol{u}^{n-1}_{t}\left(\boldsymbol{x}^{n-1}_{t}\right)\Delta t, \\
\boldsymbol{x}^{n}_{m}=\boldsymbol{x}^{n-1}_{m}+\boldsymbol{u}^{n-1}_{m}\left(\boldsymbol{x}^{n-1}_{m}\right)\Delta t,
\end{split}\end{equation}
where the superscript $n$ indicates the time level.
Defining the error as the difference between the true and modelled trajectory we find \cite{Buehner2000}
\begin{equation}
\boldsymbol{\xi}^{n}=\boldsymbol{\xi}^{n-1}+\left(\boldsymbol{u}^{n-1}_{t}(\boldsymbol{x}^{n-1}_{t})-\boldsymbol{u}^{n-1}_{m}(\boldsymbol{x}^{n-1}_{m})\right)\Delta t.
\end{equation}
We can relate the true velocity to that at the modelled particle location using the definition of the error
\begin{equation}\begin{split}
\boldsymbol{u}^{n-1}_{t}(\boldsymbol{x}^{n-1}_{t}) &=\boldsymbol{u}^{n-1}_{t}(\boldsymbol{x}^{n-1}_{m}+\boldsymbol{\xi}^{n-1})\\
                                          & \approx \boldsymbol{u}^{n-1}_{t}(x^{n-1}_{m})+\boldsymbol{\xi}^{n-1}\cdot \nabla \boldsymbol{u}^{n-1}_{t}.
\end{split}\end{equation}
Then, split the true velocity into a large and small scale component (i.e. $\bs{u}_{t}=\bs{u}_{l}+\bs{u}_{s}$), and assume the large component is equal to the modeled field to obtain (neglecting products of $\bs{\xi}$ and the small scale velocity)
\begin{equation}
\boldsymbol{\xi}^{n}=\boldsymbol{\xi}^{n-1}+\left(\boldsymbol{u}^{n-1}_{s}(\boldsymbol{x}^{n-1}_{m})+\boldsymbol{\xi}^{n-1}\cdot \nabla \boldsymbol{u}^{n-1}_{l}\right) \Delta t.
\end{equation}
This is a discrete form of the following equation, where we assume that $\bs{u}_{l} \simeq \ov{\bs{u}}$ and $\bs{u}_{s} \simeq \bs{u}'$
\begin{equation}
\frac{D\boldsymbol \xi}{Dt}=\boldsymbol{u}'+\boldsymbol{\xi}\cdot \nabla\ov{\boldsymbol{u}}.
\end{equation}
Setting $D\boldsymbol{\xi}/Dt=0$, which means the error is frozen along a particle trajectory, or that all of the fluctuation is contained in the Eulerian field \cite{Holm1999}, we arrive at the definition of the velocity fluctuation from before, equation \eqref{vfluc}. Decomposing the velocity in the Lagrangian in equation \eqref{action} into a mean and a  fluctuation, substituting the velocity fluctuation from \eqref{vfluc}, and averaging (using $\langle \boldsymbol \xi \rangle =0$, which means the error is unbiased) yields the averaged Lagrangian,
\begin{equation}
\langle L \rangle =\int_{V} \left( \frac{D}{2}\left(\ov{u}_{i}\ov{u}_{i}+\langle \xi_{k} \xi_{l}\rangle\frac{\partial \ov{u}_{i}}{\partial x_{k}}\frac{\partial \ov{u}_{i}}{\partial x_{l}}\right)+p(1-D)\right)d^{3}x.
\label{thisisit}
\end{equation}
Here we have followed the notation used in Holm \cite{Holm1999} where the averaged velocity is $\ov{\bs{u}}$, but where we keep the brackets for the averaged displacement covariance $\langle \xi_{k}\xi_{l}\rangle$. We have in equation \eqref{thisisit} essentially our turbulence model. We can see that the kinetic energy is composed of two parts, the first is the kinetic energy of the mean flow, and the second part is the  kinetic energy of the fluctuating component (or eddy kinetic energy). If we consider only the diagonal components of $\langle \xi_{k} \xi_{l} \rangle$,  the energy due to the fluctuating part will remain positive. 

\subsection{Varying the action}
To obtain our momentum equation we need to set the first variation of the action to zero. The action is defined in the usual manner
\begin{equation}
S=\int_{t_{1}}^{t_{2}}\langle L \rangle \, dt,
\end{equation}
with $\langle L \rangle$ defined in equation \eqref{thisisit}.
%To apply Hamilton's principle we consider variations with  $\ov{u}_{i}$, $D$ and $\langle \xi_{k}\xi_{l}\rangle$. 
%(see Gelfand middle of pg.25):
%\begin{equation}
%\Delta S=\int_{t_{1}}^{t_{2}}\int_{V}\left(L(\ov{u}_{i}+\delta \ov{u}_{i},\ov{u}_{i,k}+\delta \ov{u}_{i,k},D+\delta D,\langle \xi_{k}\xi_{l}\rangle+\delta \langle \xi_{k}\xi_{l}\rangle)-L(\ov{u}_{i},\ov{u}_{i,k},D,\langle \xi_{k} \xi_{l} \rangle )\right)d^{3}\boldsymbol{x} \, dt,
%\end{equation}
The first variation of the action is
\begin{equation}
\delta S=\int_{t_{1}}^{t_{2}}\int_{V}\left(\frac{\partial \langle l \rangle}{\partial \ov{u}_{i}} \delta \ov{u}_{i}+\frac{\partial \langle l \rangle}{\partial D}\delta D+ \frac{\partial \langle l \rangle}{\partial \langle \xi_{k}\xi_{l} \rangle}\delta \langle \xi_{k}\xi_{l} \rangle \right) d^{3}x \, dt,
\label{varaction}
\end{equation}
where $(l)$ is the Lagrangian density (Lagrangian/unit volume). The partial derivatives with respect to the volume element and particle displacement are
\begin{align}
\frac{\partial \langle l \rangle}{\partial D} &= \frac{\ov{u}_{i}\ov{u}_{i}}{2}+\frac{\langle \xi_{k}\xi_{l}\rangle}{2} \frac{\partial \ov{u}_{i}}{\partial x_{k}}\frac{\partial \ov{u}_{i}}{\partial x_{l}}-p,
\label{Dderiv}\\
\frac{\partial \langle l \rangle}{\partial \langle \xi_{k}\xi_{l}\rangle} &= 
\frac{D}{2}\frac{\partial \ov{u}_{i}}{\partial x_{k}}\frac{\partial \ov{u}_{i}}{\partial x_{l}}.
\label{covarderiv}
\end{align}
For the velocity
\begin{align}
\int_{V} \frac{\partial \langle l \rangle}{\partial \ov{u}_{i}} \delta \ov{u}_{i} \, dV=& \int_{V} D\ov{u}_{i}\delta \, \ov{u}_{i}+D\frac{\langle \xi_{k}\xi_{l}\rangle}{2}
\left(
  \frac{\partial}{\partial x_{k}}\left(\delta \ov{u}_{i}\right)\frac{\partial \ov{u}_{i}}{\partial x_{l}}
+\frac{\partial \ov{u}_{i}}{\partial x_{k}}\frac{\partial}{\partial x_{l}}\left(\delta \ov{u}_{i}\right)
\right) \, d^{3}x \\
=& \int_{V} D\ov{u}_{i}\delta \ov{u}_{i}+D\langle \xi_{k}\xi_{l}\rangle
\left(\frac{\partial \ov{u}_{i}}{\partial x_{l}}\frac{\partial}{\partial x_{k}}\left(\delta \ov{u}_{i}\right)\right) \, d^{3}x \\
%\end{align}
%Integrating by parts
%\begin{align}
%\int_{V} \frac{\partial \langle l \rangle}{\partial \ov{u}_{i}} \delta \ov{u}_{i} \, d^{3}x
%=& \int_{V} D\ov{u}_{i}\delta \ov{u}_{i}+D\langle \xi_{k}\xi_{l}\rangle
%\left(\frac{\partial \ov{u}_{i}}{\partial x_{l}}\frac{\partial}{\partial x_{k}}\left(\delta \ov{u}_{i}\right)\right)\,d^{3}x,\\
=&
\int_{A}D\langle \xi_{k}\xi_{l}\rangle \frac{\partial \ov{u}_{i}}{\partial x_{l}} \delta \ov{u}_{i}\,dA_{k}
+\int_{V}\left(D\ov{u}_{i}-\frac{\partial}{\partial x_{k}}\left( D\langle \xi_{k}\xi_{l}\rangle \frac{\partial \ov{u}_{i}}{\partial x_{l}}\right) \right) \delta \ov{u}_{i} \,d^{3}x,
\label{lvaru}
\end{align}
where we have applied symmetry of the particle displacement covariance $\langle \xi_{k}\xi_{l} \rangle$, and applied integration by parts in the last step. If we apply either a periodic boundary condition or the constraint that the normal component of $\langle \xi_{k} \xi_{l}\rangle$ is zero, the surface integral in \eqref{lvaru} is zero and we  have
\begin{equation}
\int_{V} \frac{\partial \langle l \rangle}{\partial \ov{u}_{i}} \delta \ov{u}_{i} \, d^{3}x
=\int_{V} \left( D\ov{u}_{i}-\frac{\partial}{\partial x_{k}}\left( D\langle \xi_{k}\xi_{l}\rangle \frac{\partial \ov{u}_{i}}{\partial x_{l}}\right) \right) \delta \ov{u}_{i} \, d^{3}x.
\label{uderiv}
\end{equation}

\subsection{Definition of the variations of Eulerian coordinates}
We have now defined the partial derivatives which appear in our varied action \eqref{varaction}, it still remains to define the variations of the Eulerian coordinates $\delta \ov{u}_{i}, \delta D$ and $\delta \langle \xi_{k}\xi_{l} \rangle$. If we were working with the material representation, given by equation \eqref{material}, variations would have been taken with respect to the particle position, and all that would remain to be done would be to set the first variation to zero.  In the spatial picture we have followed here variations of the Eulerian coordinates ($\boldsymbol{u},\rho,s$) are taken at  a fixed point, although the final goal is the same as in the material representation, to find the trajectory for which the action is stationary. The first question is then, how are variations of a particle trajectory reflected in the Eulerian coordinates? 
%Consider that if we are sitting at a fixed point watching particles being advected past, we have different particles passing through our fixed point. We can anticipate that the variation of an Eulerian variable will need to consider the variation of a particle trajectory (as seen from a fixed position), and the fact that different labels will occupy that position.  
\\
\\
To connect the two we define a function that relates the field position $\boldsymbol{x}$ to a label $\boldsymbol{a}$. Such a function is given by the trajectory, which we will denote by $\boldsymbol{\eta}$. More formally,  let $\boldsymbol{\eta}$ be the function that maps particles with labels $\boldsymbol{a}$ to the field points they occupy at time $t$. For our purposes here we will assume this map is one-to-one, invertible and sufficiently smooth that we may differentiate it as many times as necessary. The particle position $\boldsymbol{x}$ is 
\begin{equation}
\boldsymbol{x}=\boldsymbol{\eta}(\boldsymbol{a},t).
\end{equation}
Similarly, let $\boldsymbol{\eta}^{-1}$ be the map that tells you the label $\bs{a}$ of the particle occupying the field point $\bs{x}$ at time t,
\begin{equation}
\boldsymbol{a}=\boldsymbol{\eta}^{-1}(\boldsymbol{x},t).
\end{equation}
The Eulerian and Lagrangian velocities at a given point are related through the identity
\begin{equation}
\bs{u}=\dot{\bs{\eta}}(\bs{\eta}^{-1}(\bs{x},t),t),
\end{equation}
or
\begin{equation}
\boldsymbol{u}=\dot{\boldsymbol{\eta}} \circ \boldsymbol{\eta}^{-1}
\label{urelation}
\end{equation}
where the $\circ$ operator denotes a composition of maps \cite{Spivak1994}, and the dot indicates a time derivative. Consider that if we only know the mapping $\bs{\eta}$, then to find the velocity, $\boldsymbol{u}(\boldsymbol{x},t)$, at a given field point $\boldsymbol{x}$, we can evaluate $\boldsymbol{a}=\boldsymbol{\eta}^{-1}(\boldsymbol{x},t)$ at our field point to find the particle occupying that point at time $t$. Knowing the particle (denoted by the label $\boldsymbol{a}$) we can then evaluate $\dot{\boldsymbol{\eta}}(\boldsymbol{a},t)$ to get the velocity at that point. The central idea to keep in mind here is that in the Eulerian framework the dependence of the velocity field on the particle trajectory comes into play in two places, one in calculating the rate of change of the trajectory and the other in evaluating the label. This means when we vary the velocity field we need to take both of these into account. 
\\
\\
The trajectory is related to the volume element according to
\begin{equation}
D(\bs{x},t)=\frac{1}{det(\nabla \bs{\eta})}.
\label{Drelation}
\end{equation}
Finally, the  displacement covariance equation, 
\begin{equation}\frac{D \langle \xi_{k}\xi_{l} \rangle}{Dt}=0, \end{equation}
can be written in material form as
\begin{equation}
\langle \xi_{k}\xi_{l} \rangle \circ \bs{\eta} = {\langle \xi_{k}\xi_{l} \rangle}_{o}
\label{advectmat}
\end{equation}
where the subscript $o$ denotes the initial value. This equation tells us the displacement covariance at the current particle location $\bs{\eta}$ is equal to the initial displacement covariance, in other words that the displacement covariance is preserved along trajectories. 
From these expressions for the velocity \eqref{urelation}, volume element \eqref{Drelation}, and displacement covariance \eqref{advectmat} we will be able to relate variations of the Eulerian coordinates to the particle trajectory.
\\
\\
As an example consider the variation of the displacement covariance. We start with the definition of the particle trajectory variation \cite{Bhat2005}
\begin{equation}
\delta \boldsymbol{\eta} := \left. \frac{d}{d\epsilon} \right |_{\epsilon=0}  \underbrace{\left (\boldsymbol{\eta}+\epsilon \, \delta \boldsymbol{\eta} \right)}_{\boldsymbol{\eta}^{\epsilon}}.
\label{trajvar}
\end{equation}
Variations of the Eulerian coordinates, $\bs{u},D$ and $\langle \xi_{k}\xi_{l} \rangle$, are defined in a similar way. Now vary both sides of \eqref{advectmat} and differentiate with respect to $\epsilon$ noting that the RHS is constant
\begin{equation}
\left. \frac{d}{d\epsilon} \right |_{\epsilon=0} \left(\langle \xi_{k}\xi_{l} \rangle^{\epsilon}\circ \bs{\eta}^{\epsilon}\right)=0.
\end{equation}
Denoting the varied quantities as 
%\begin{equation}
\begin{align}
\langle \xi_{k}\xi_{l}\rangle^{\epsilon}&=\langle \xi_{k}\xi_{l}\rangle + \epsilon \delta \langle \xi_{k} \xi_{l} \rangle\\
\bs{\eta}^{\epsilon}&=\bs{\eta}+\epsilon \delta \bs{\eta}
\end{align}
%\end{equation}
 and carrying out the differentiation gives
\begin{equation}
\nabla \langle \xi_{k}\xi_{l}\rangle \cdot \delta \bs{\eta}+\delta \langle \xi_{k}\xi_{l} \rangle  \circ \bs{\eta}=0.
\end{equation}
Composing both sides with $\bs{\eta}^{-1}$
\begin{equation}
\delta \langle \xi_{k}\xi_{l}\rangle=-\nabla \langle \xi_{k}\xi_{l}\rangle \cdot \bs{w}
\label{Dispvar}
\end{equation}
where $\boldsymbol{w}$ is the trajectory variation expressed at a field point, 
\begin{equation}
\boldsymbol{w}=\delta \boldsymbol{\eta} \circ \boldsymbol{\eta}^{-1}. 
\end{equation}
In a similar manner it can be shown that the velocity and volume element variations are  \cite{Scott2008,Bhat2005}
\begin{equation}
\delta \boldsymbol{\ov{u}}=\frac{\partial \boldsymbol{w}}{\partial t}+\boldsymbol{\ov{u}}\cdot \nabla \boldsymbol{w}-\boldsymbol{w}\cdot \nabla \boldsymbol{\ov{u}},
\label{uvar1}
\end{equation}
\begin{equation}
\delta D =- \nabla \cdot \left(D \bs{w} \right).
\label{Dvar}
\end{equation}
The first two terms in the velocity variation are due to the rate of change of the trajectory variation, with the label fixed, while the last term is due to the variation with respect to the label \cite{Scott2008,Bhat2005}. This is what we meant earlier when we said earlier that both the trajectory and label need to be varied. Expanding the variation of the volume element gives two terms that can be interpreted similarly. A more detailed discussion on the variation of Eulerian quantities can be found in Bretherton \cite{Bretherton1970}.  

\subsection{Setting the first variation to zero} 
To proceed with setting the first variation of the action, given by equation \eqref{varaction} to zero, we substitute the variations given by equations\eqref{uvar1},\eqref{Dvar},\eqref{Dispvar} into the varied action \eqref{varaction},
\begin{equation}
\delta S =
\left(
\int_{t_{1}}^{t_{2}}\int_{V}
\frac{\partial \langle l \rangle}{\partial u_{i}}
\left(\frac{\partial w_{i}}{\partial t}+\ov{u}_{j}\frac{\partial w_{i}}{\partial x_{j}}-w_{j}\frac{\partial \ov{u}_{i}}{\partial x_{j}} \right)
-\frac{\partial}{\partial x_{i}}(D w_{i})\frac{\partial \langle l \rangle }{\partial D} 
-\frac{\partial \langle l \rangle}{\partial \langle \xi_{k}\xi_{l} \rangle}\frac{\partial \langle \xi_{k}\xi_{l}\rangle}{\partial x_{i}}w_{i}\right)
d^{3}x \, dt.
\label{deltaS2}
\end{equation}
Integrating by parts and changing sign
\begin{equation}\begin{split}
\delta S =&\int_{t_{1}}^{t_{2}}\int_{V}
\left[
\left(\frac{\partial}{\partial t}+\ov{u}_{j}\frac{\partial}{\partial x_{j}}\right)
\frac{\partial \langle l \rangle}{\partial \ov{u}_{i}}
+\frac{\partial \langle l \rangle}{\partial \ov{u}_{j}}\frac{\partial \ov{u}_{j}}{\partial x_{i}}
-D\frac{\partial}{\partial x_{i}}\frac{\partial \langle l \rangle}{\partial D}
+\frac{\partial \langle l \rangle}{\partial \langle \xi_{k}\xi_{l} \rangle}\frac{\partial \langle \xi_{k}\xi_{l}\rangle}{\partial x_{i}}\right]
w_{i}
d^{3}x \, dt\\
-&\underbrace{\int_{V} \left. \frac{\partial \langle l \rangle}{\partial \ov{u}_{i}}w_{i} \,dV \right |_{t_{1}}^{t_{2}}}_{I}
-\underbrace{\int_{t_{1}}^{t_{2}}\int_{A}\frac{\partial \langle l \rangle}{\partial \ov{u}_{i}} \ov{u}_{j}w_{i}\,dA_{j}}_{II}
+\underbrace{\int_{t_{1}}^{t_{2}}\int_{A}\frac{\partial \langle l \rangle}{\partial D} D w_{i}\,dA_{i}}_{III}
\label{varLag}
\end{split}\end{equation}
The last three terms are zero for the following reasons: \\
(I) variations are zero at beginning and end times (same as for Newton's law), \\
(II) velocity is either periodic or has zero normal component for a solid surface, \\
(III) trajectory variation $(\bs{w})$ is tangent to the bounding surface \cite{Morrison1998}.
\\
\\
Substituting in the partial derivatives \eqref{uderiv}, \eqref{Dderiv}, \eqref{covarderiv} into \eqref{deltaS2}, taking the limit as $d^{3}x$,$w_{i}$, and $dt$ go to zero (c.f. Gelfand \cite{Gelfand1963}) and applying the constraint $D=1$, yields the momentum equation
\begin{equation}
\frac{\partial v_{i}}{\partial t}+\ov{u}_{j}\frac{\partial v_{i}}{\partial x_{j}}+v_{j}\frac{\partial \ov{u}_{j}}{\partial x_{i}}=-\frac{\partial p^{\alpha}}{\partial x_{i}}-\frac{1}{2}\frac{\partial \langle \xi_{k}\xi_{l}\rangle}{\partial x_{i}}\frac{\partial \ov{u}_{m}}{\partial x_{k}}\frac{\partial \ov{u}_{m}}{\partial x_{l}}
\label{Holm280}
\end{equation}
where the following variables have been defined
\begin{align}
& v_{i}=\ov{u}_{i}-\frac{\partial}{\partial x_{k}}\left(\langle \xi_{k}\xi_{l}\rangle \frac{\partial \ov{u}_{i}}{\partial x_{l}}\right) \\
& p^{\alpha}=p-\frac{1}{2}\ov{u}_{m}\ov{u}_{m}-\langle \xi_{k}\xi_{l} \rangle \frac{\partial \ov{u}_{i}}{\partial x_{k}}\frac{\partial \ov{u}_{i}}{\partial x_{l}}.
\end{align}
Note that there are two velocities, related through the Helmholtz operator,
\begin{equation}
v_{i}=\underbrace{\left(1-\frac{\partial}{\partial x_{k}}\left(\langle \xi_{k}\xi_{l} \rangle \frac{\partial}{\partial x_{l}}\right)\right)}_{H}\ov{u}_{i}
\label{Helm}
\end{equation}
%point out here that the definition of the unsmoothed velocity arises from the smoothed instead of vice-versa - as is the usual case in LES
If we impose isotropy $\langle \xi_{k}\xi_{l}\rangle=\alpha^{2}\delta_{kl}$ and assume $\alpha^{2}$ is constant, we arrive at the inviscid form of the NS-$\alpha$ equations found in the literature \cite{Holm2003,Geurts2006}
\begin{equation}
\frac{\partial v_{i}}{\partial t}+\ov{u}_{j}\frac{\partial v_{i}}{\partial x_{j}}+v_{j}\frac{\partial \ov{u}_{j}}{\partial x_{i}}=-\frac{\partial p^{\alpha}}{\partial x_{i}},
\end{equation}
with
\begin{equation}
v_{i}=\ov{u}_{i}-\alpha^{2}\frac{\partial^{2} \ov{u}_{i}}{\partial x^{2}_{k}}.
\end{equation}
In Fourier space 
\begin{equation}
\hat{\tl{u}}_{i}(\bs{k})=\frac{\hat{v}_{i}(\bs{k})}{1+\alpha^{2}|\bs{k}|^{2}},
\end{equation}
from which it is clear that the smoothed velocity is low-pass filtered since the high-wavenumber components are attenuated. From this relationship we can see that $\alpha$ can be interpreted as a filter width. 
\\
\\
The continuity equation does not come from the variational principle, but from taking the time derivative of the volume element \cite{Holm1999}
\begin{equation}
\frac{\partial D}{\partial t}+\nabla \cdot (D\ov{\bs{u}})=0.
\end{equation}
Imposing the constraint $D=1$ then gives $\nabla \cdot \ov{\bs{u}}=0$, or $\nabla \cdot \tl{\bs{u}}=0$ when we recognize $\ov{\bs{u}}$ is the smoothed velocity according to the Helmholtz operation.
\\
\\
There are a few things to note here. The first is that the averaged velocity $\bs{\ov{u}}$ becomes the smoothed velocity when we consider that $\bs{v}$ and $\bs{\ov{u}}$ are related through a Helmholtz operator. This is what is meant in the literature by `temporal averaging in the variational principle implies a spatial smoothing in the momentum equation' \cite{Holm1999}, although one could anticipate this from the expression for the velocity fluctuation, which was derived using a spatial Taylor series expansion \eqref{tsvfluc}. Another aspect to note is that if we had not considered the functional dependence on $\langle \xi_{k} \xi_{l}\rangle$ we would not have obtained the $\partial \langle \xi_{k}\xi_{l}\rangle/\partial x_{i}$ term in the final momentum equation \eqref{Holm280}. This term is necessary to conserve momentum, which you can see either by considering Noether's theorem for the action principle, or by removing this term and trying to write the momentum equation in conservative form.  Some studies of the NS-$\alpha$ equations have not included this term in their analyses yielding incorrect results, as pointed out in the literature \cite{Putkaradze2003,Holm2003b}. Finally, by following through with this method we are able to understand how the boundary conditions for the NS-$\alpha$ equations arise.  
\\
\\
Note that other methods can also be used to derive the equations from Hamilton's principle \cite{Holm1999, Marsden2003, Montgomery2002,Chen1999c}. The method given here is both straightforward and general enough to allow the model parameter $\langle \xi_{k}\xi_{l} \rangle$ to be non-constant and anisotropic. Extensions of the model to stratified and rotating flows are given by Holm \cite{Holm1998,Holm1999}, and can be obtained using the methods used here. For alternative examples on the use of variational principles in fluid mechanics see Salmon \cite{Salmon1998} and Finlayson \cite{Finlayson1972}. When using their methods the advection equation, $D\langle \xi_{k} \xi_{l} \rangle /Dt=0$, needs to be added as a constraint equation to obtain the last term on the RHS of Equation \eqref{Holm280} that contains the gradient of the particle displacement covariance. 

\section{Demonstration}

\subsection{Model Formulation}
\label{formulation}
The formulation of the NS-$\alpha$ model has been described in \cite{Scott2010} and is briefly reviewed here for continuity. To investigate the NS-$\alpha$ model numerically we work within an LES template and develop an equation with the smoothed velocity as the dependent variable. To do this, first add a viscous term to \eqref{Holm280} and then rewrite the equation in momentum-conservation form \cite{Chen1999b} (now replacing $\ov{\bs{u}}$ with $\tl{\bs{u}}$)
\begin{equation}
\frac{\partial v_{i}}{\partial t}+\tl{u}_{j}\frac{\partial v_{i}}{\partial x_{j}}=-\frac{\partial p}{\partial x_{i}} + \frac{\partial}{\partial x_{j}}\left(\langle \xi_{k}\xi_{j} \rangle 
\frac{\partial \tl{u}_{m}}{\partial x_{i}}\frac{\partial \tl{u}_{m}}{\partial x_{k}} \right)+\nu\frac{\partial^{2} v_{i}}{\partial x_{k}^{2}}.
\label{NSalpha_anis}
\end{equation}
Then, to write the substantial derivative entirely in terms of the smoothed velocity, rewrite the advective terms in  \eqref{NSalpha_anis} as,
\begin{equation}
\frac{\partial v_{i}}{\partial {t}}+\tl{u}_{j}\frac{\partial v_{i}}{\partial x_{j}}=[D/Dt,H]\tl{u}_{i}+H\left(\frac{\partial \tl{u}_{i}}{\partial t}+\tl{u}_{j}\frac{\partial \tl{u}_{i}}{\partial x_{j}} \right).
\label{advective}
\end{equation}
Here $[D/Dt,H]$ is the commutator between the material derivative and the Helmholtz operator, $H$ from equation \eqref{Helm}, 
\begin{equation}
[D/Dt,H]\tl{u}_{i}=D/Dt(H(\tl{u}_{i}))-H(D/Dt(\tl{u}_{i})), 
\end{equation}
where $H(\tl{u}_{i})=v_{i}$. Note that the substantial derivative is defined with the smoothed velocity,  $D/Dt=\partial_{t}+\tl{u}_{j}\partial_{j}$. The momentum equation \eqref{NSalpha_anis} can then be written as,
\begin{equation}
\frac{\partial \tl{u}_{i}}{\partial t}+\tl{u}_{j}\frac{\partial \tl{u}_{i}}{\partial x_{j}}=H^{-1}\left(
-\frac{\partial p}{\partial x_{i}} + \frac{\partial}{\partial x_{j}}\left(\langle \xi_{k}\xi_{j} \rangle 
\frac{\partial \tl{u}_{m}}{\partial x_{i}}\frac{\partial \tl{u}_{m}}{\partial x_{k}} \right)+\nu\frac{\partial^{2} v_{i}}{\partial x_{k}^{2}}-[D/Dt,H]\tl{u}_{i}\right).
\end{equation}
Expanding the commutator, applying $D\langle \xi_{k}\xi_{l} \rangle/Dt=0$, the momentum equation can then be written
\begin{equation}
\frac{\partial \tl{u}_{i}}{\partial t}+\tl{u}_{j}\frac{\partial \tl{u}_{i}}{\partial x_{j}}=
-\frac{\partial \tl{p}}{\partial x_{i}}+\nu\frac{\partial^{2} \tl{u}_{i}}{\partial x_{k}^{2}}
-H^{-1}\left(\frac{\partial m_{ij}}{\partial x_{j}}\right).
\end{equation}
The subgrid stress is
\begin{equation}\begin{split}
 m_{ij}
=\langle \xi_{k}\xi_{l} \rangle \frac{\partial \tl{u}_{i}}{\partial x_{k}}\frac{\partial \tl{u}_{j}}{\partial x_{l}}
+\langle \xi_{j}\xi_{l} \rangle \frac{\partial \tl{u}_{k}}{\partial x_{l}}\frac{\partial \tl{u}_{i}}{\partial x_{k}}
-\langle \xi_{k}\xi_{j}\rangle \frac{\partial \tl{u}_{m}}{\partial x_{i}}\frac{\partial \tl{u}_{m}}{\partial x_{k}}.
\label{mijfull}
\end{split}\end{equation}
Instead of using the full anisotropic model, in the following demonstration we consider a simplified version that arises when only the diagonal components of $\langle \xi_{k}\xi_{l}\rangle$ are retained, which ensures that the kinetic energy in the Lagrangian (equation \eqref{thisisit}) is positive. Denoting $\alpha^{2}_{k}=\langle \xi_{k}\xi_{k}\rangle$ we arrive at our subgrid stress 
\begin{equation}
m_{ij}=\underbrace{
\alpha^{2}_{k}\delta_{kl} \frac{\partial \tl{u}_{i}}{\partial x_{k}}\frac{\partial \tl{u}_{j}}{\partial x_{l}}}_{A_{ij}}
+\underbrace{\alpha^{2}_{l}\delta_{jl} \frac{\partial \tl{u}_{k}}{\partial x_{l}}\frac{\partial \tl{u}_{i}}{\partial x_{k}}}_{B_{ij}}
-\underbrace{\alpha^{2}_{k}\delta_{kj} \frac{\partial \tl{u}_{m}}{\partial x_{i}}\frac{\partial \tl{u}_{m}}{\partial x_{k}}}_{C_{ij}}.
\label{sgsstress}
\end{equation}
This is the subgrid stress that would result if the Helmholtz operator is considered as being equivalent to the composition of three one-dimensional, symmetric filters (the off-diagonal components now being zero). A similar filter has been used in the Tensor-Diffusivity model \cite{Winckelmans2001}. Here it reduces the cost of the model such that,  when the explicit filter is applied by solving the Helmholtz equation using Fourier transforms, the model adds approximately $30\%$ to the total computational time, similar to what is found in other studies that used a constant, isotropic model parameter, $\alpha^{2}\delta_{kl}=\langle \xi_{k}\xi_{l} \rangle$ \cite{Hecht2008}.

\subsection{Description of the test Case}
The test case chosen here is turbulent channel flow. The focus here is on wall-resolving LES, thus we are going to consider Reynolds numbers at the low end of the turbulent regime. This is a challenging test case for the present model because it is a model with a modified nonlinearity, while in this test case diffusion plays a prominent role.  Channel flow with Reynolds number $Re_{\tau}=180$ is studied here using  a second-order finite volume method \cite{Lien1994a}. Periodic boundary conditions are applied in the homogeneous directions (streamwise and spanwise) while no-slip conditions are used for the solid boundaries located at $y= \pm H$, where $H$ is the channel half-height. The mesh is uniformly spaced in the homogeneous directions but stretched in the wall-normal directions using a hyperbolic tangent profile. To enable a variety of subgrid and numerical resolutions to be tested, the investigation was done primarily using the minimal channel flow \cite{Jimenez1991}. This is the smallest domain for which the near wall cycle is able to sustain turbulence. The near wall cycle consists of interactions between low-speed spanwise streaks, streamwise vortices, and hairpin vortices \cite{Robinson1991}. To sustain turbulence, the channel must be wide enough to contain a low-speed streak. For this purpose we chose a channel of dimension ($\pi, 2, 0.3\pi)$, where the non-dimensionalization is with respect to the channel half height. Various mesh resolutions were tested, summarized in Table \ref{table_minimal}. In our simulations a constant mean mass flux was enforced at a Reynolds number of $4160$ based on the centerline velocity of a laminar flow and the channel half-height $H$. This is equivalent to a bulk flow Reynolds number of $Re_{b}=2773$ or $Re_{\tau}=180$.
\\
\\
The flow was initialized using a parabolic profile with a superimposed Tollmien-Schlicting (T-S) wave to provide a 2D disturbance. A T-S wave with amplitude of $10\%$ of the centerline velocity and wavenumber 2 (made dimensionless with the channel half-height) was found to bring the flow to a turbulent state quickly. This method was preferred over white noise because it was found the noise had a tendency to require a longer time to reach a turbulent state.  By initializing the flow with a large scale disturbance, nonlinear interactions quickly generate a cascade of energy towards the small scales. 
\\
\\
In the results an averaged quantity is denoted by an overbar, and a fluctuation about this state is denoted with a prime. For the velocity, vorticity and other profiles reported (quantities that are a function of the vertical coordinate) the averaging is taken over the statistically homogeneous directions $x$ and $z$ as well as with time to increase the statistical sample. Quantities are non-dimensionalized using the channel half-height, $H$, and the shear velocity $u_{\tau} \equiv \sqrt{\tau_{w}}$, where $\tau_{w}$ is the wall shear stress
$\tau_{w} \equiv \left . \nu \frac{\partial u}{\partial y}\right |_{w}$. These non-dimensional quantities are $u^{+}=u/u_{\tau}$, $y^{+}=\frac{y u_{\tau}}{\nu}$, $Re_{\tau}=\frac{u_{\tau} H}{\nu}$.

\subsection{Definition of $\alpha^{2}_{k}$}
\label{alphadef}
To specify $\alpha^{2}_{k}$ here as a first step we followed a conventional LES approach and based $\alpha^{2}_{k}$ on the mesh spacing
\begin{equation}
\alpha^{2}_{k}=C\left(h_{k}^{2}\right)
\label{alphamesh}
\end{equation}
where $h_{k}$ is the grid spacing in the \textit{k}-direction and $C$ is a constant denoting what fraction of the grid spacing to use.  Because $\alpha_{k}^{2}$ can be related to the width, $\Delta_{k}$, of a box filter via $\alpha_{k}^{2}=\Delta^{2}_{k}/24$ \cite{Geurts2003}, we choose $C=1/6$, which corresponds to a filter width which is twice the grid size. To avoid the commutation error that arises when a filter with non-uniform widths is used, we chose not to filter in the wall-normal direction, and for this reason $\alpha^{2}_{y}$ was set to zero. The filter was applied both by solving the Helmholtz equation using Fourier transforms and also by using a box filter. Results using the two methods were very similar \cite{Scott2008}, those shown here solve the Helmholtz equation. It should also be noted that we also tried using the isotropic model with $\alpha^{2}$ based on the grid volume, but were not able to achieve numerically stable results with that definition of $\alpha^{2}$. 
\subsection{Results}
\label{results}
\subsubsection{Mean flow and energy transfer}
\begin{figure}
\psfrag{uplus}{$u^{+}$}
\psfrag{yplus}{$y^{+}$}
\mbox{
%\subfigure[Mean flow]{\includegraphics[width=8cm,height=5.6cm]{../thesis/figs/channel/minimal/ulog_4grids.eps}}}
\subfigure[Mean flow]{\includegraphics[width=13cm,height=8cm]{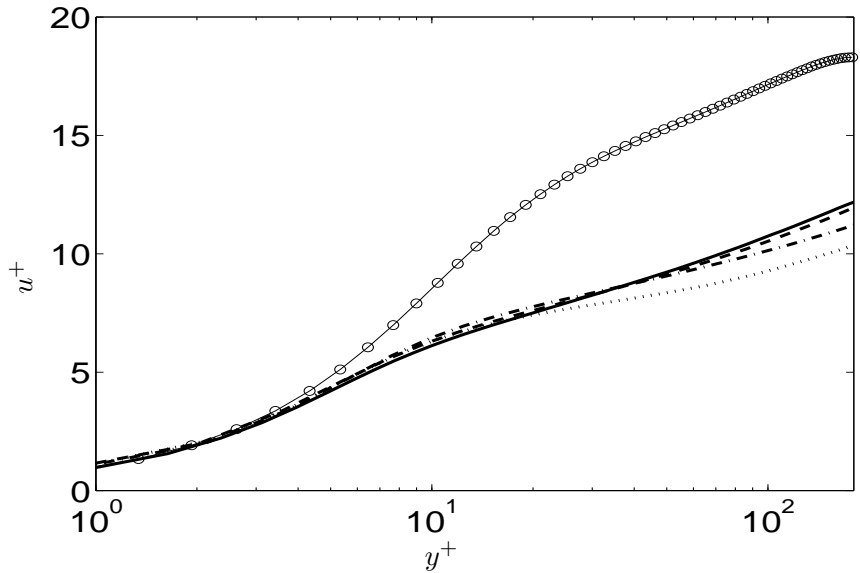}}}
\caption{Mean flow profiles for different meshes with different values of the smoothing length scale $\alpha$, in terms of parameter $C$ from equation \eqref{alphamesh}. Dash-dotted line, $C=1/6$, $(16,64,16)$; dashed line,  $C=1/6$, $(24,64,24)$; dotted line, $C=2/3$, $(32,64,32)$; solid $C=1/6$, $(32,128,32)$; symbols full channel DNS \cite{Kim1987}. To look at the effect of refining the mesh while keeping the subgrid resolution constant the dash-dotted, dashed and solid lines can be compared. To look at the effect of keeping the physical size of $\alpha^{2}_{k}$ constant while refining the mesh (hence increasing the subgrid resolution) the dash-dotted and dotted lines can be compared. }
\label{minulog}
\end{figure}

\begin{figure}[tbp]
\mbox{
\subfigure[DNS]{\includegraphics[width=7cm,height=5cm]{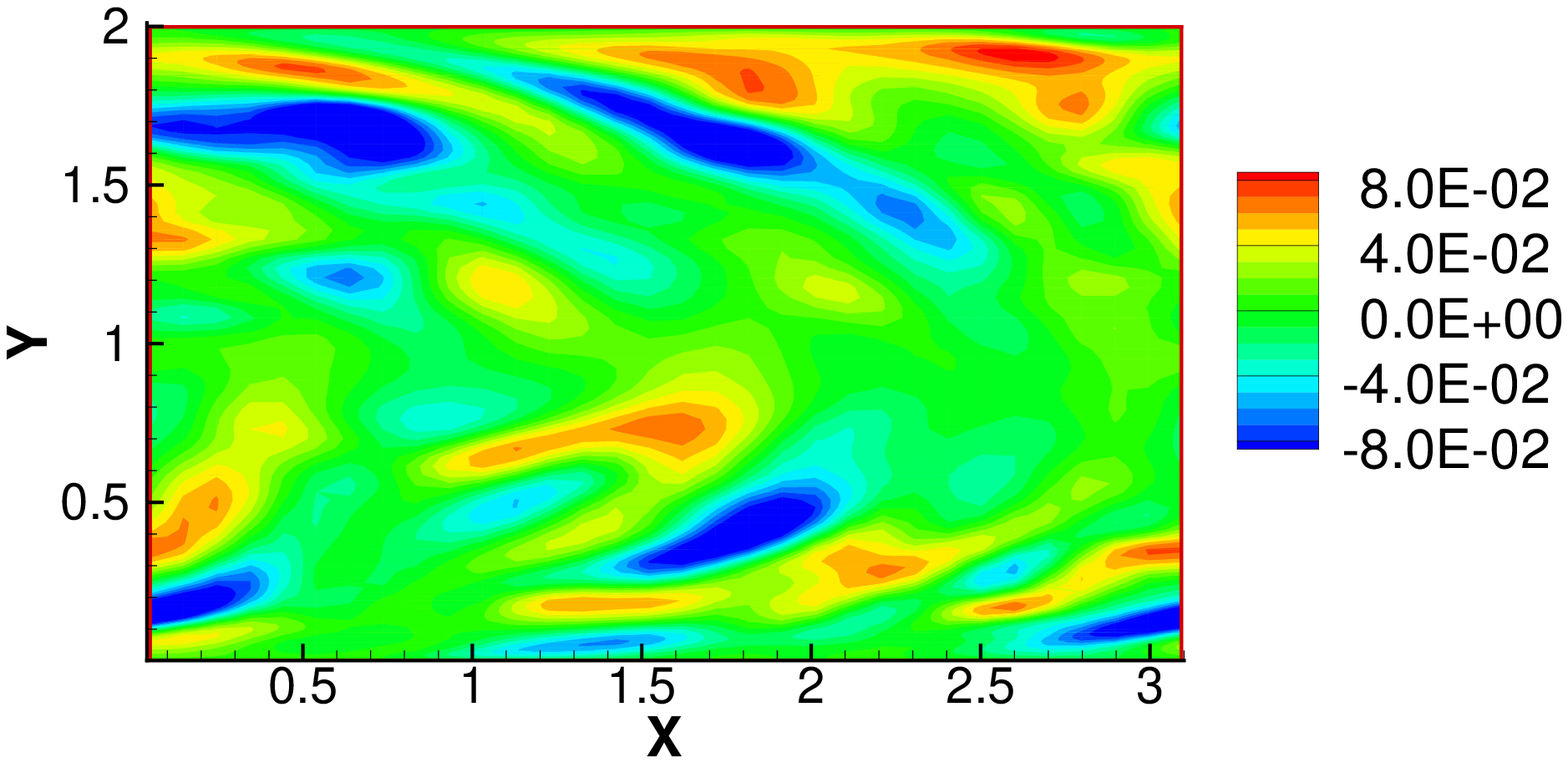}}}
\mbox{
\subfigure[NS-$\alpha$]{\includegraphics[width=7cm,height=5cm]{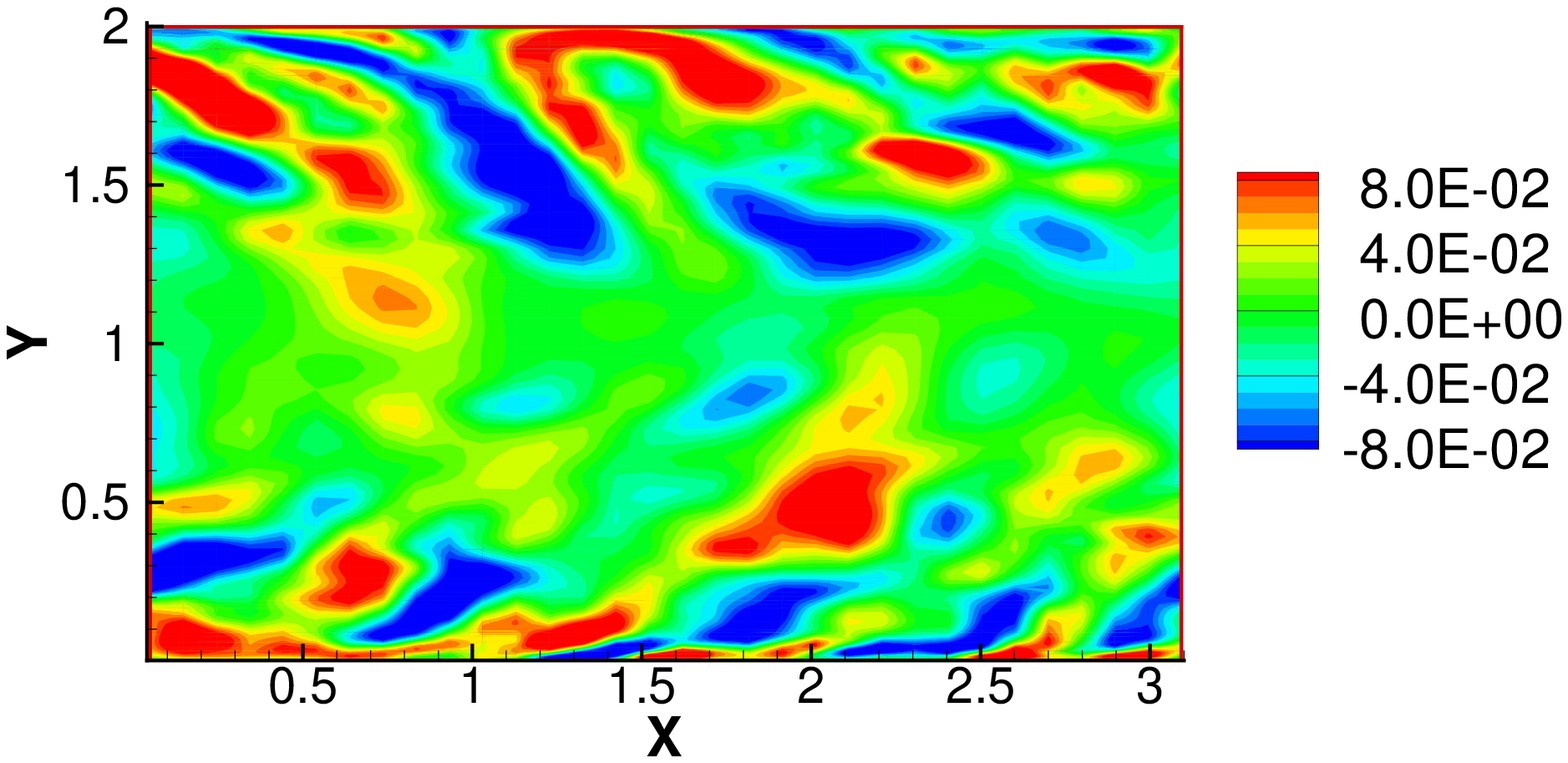}}}
\caption{Contours of the instantaneous spanwise velocity for the minimal channel for the $(32,128,32)$ mesh. There is more small scale activity with the NS-$\alpha$ model.}
\label{winst_min}
\end{figure}

\begin{figure}[tbp]
\psfrag{tsgs}{$T_{SGS}^{+}$}
\psfrag{yplus}{$y^{+}$}
\includegraphics[height=8cm]{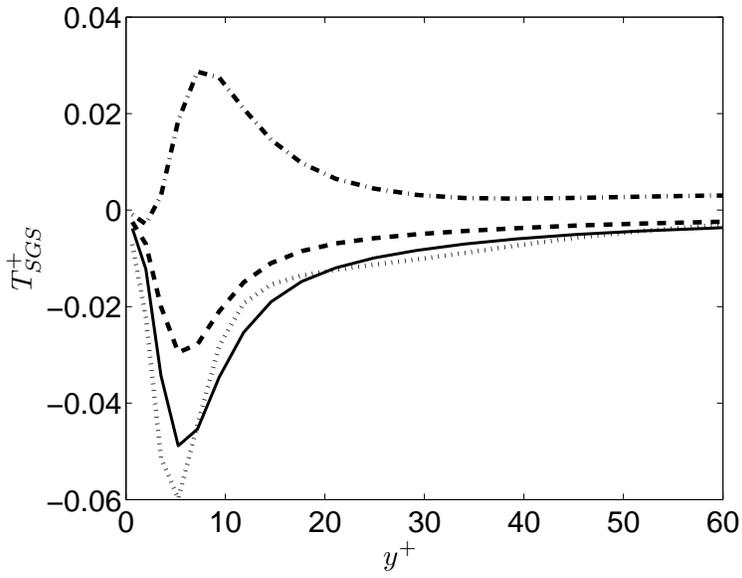}
\caption{Energy transfer $T_{SGS}^{+}=m_{ij} \partial_{j} \tl{u}_{i}$. Dashed line, $T_{SGSA}^{+}$; solid line, $T_{SGSB}^{+}$; dash-dotted line $T_{SGSC}^{+}$; dotted line, total transfer, $T_{SGSA}+T_{SGSB}-T_{SGSC}$. See equation \eqref{sgsstress} for the definitions of the $A,B$ and $C$ terms. The peak of the total energy transfer here is correct in magnitude, but too close to the wall.}
\label{TSGSprofiles}
\end{figure}

\noindent
The first quantity that is of interest in turbulent channel flow is the mean flow profile,  which is related to skin friction. The mean flow profile using the definition of $\alpha^{2}_{k}$ given in section \ref{alphadef} is shown in Figure \ref{minulog} for the four different meshes listed in Table \ref{table_minimal}. We can see the velocity is significantly underestimated on all three meshes $(16, 64, 16; 24, 64, 24; 32, 128, 32)$ using this definition of $\alpha^{2}_{k}$ with $C=1/6$. This implies that the skin friction is significantly overpredicted. To check the robustness of this result we also looked at the effect of refining the mesh while keeping the physical size of $\alpha^{2}_{k}$ constant. By using a $(32,64,32)$ mesh with $C=2/3$ in comparison with a $(16, 64, 16)$ mesh with $C=1/6$, we are able to check the effect of increasing the subgrid resolution (the ratio between the filter width and the mesh spacing). Geurts and Holm \cite{Geurts2006} found for a temporally evolving mixing layer when increasing the subgrid resolution  from $C=1/6$ to $C=1$ they were able to reduce the turbulent kinetic energy and bring their simulation results into good agreement with DNS data. It can be seen in Figure \ref{minulog} that in our case increasing the subgrid resolution does not improve the mean flow profile. We also found the spanwise velocity to be significantly overpredicted close to the wall (profiles not shown). A similar result was found in Zhao and Mohseni \cite{Zhao2005} in their study of a channel flow. We can see in Figure \ref{winst_min} that the spanwise velocity for the NS-$\alpha$ contains more small-scale activity as compared to the case with no model, and indicates coherent structures close to the grid scale. This is consistent with what is found in other studies \cite{Hecht2008b,Geurts2006,Geurts2008}. 
\\
\\
Because the NS-$\alpha$ model is non-dissipative, and also contains the backscatter dynamics, it is possible that if $\alpha^{2}_{k}$ is chosen to be too close to the energy containing scales this could lead to a build-up of energy. Here we look at the subgrid energy transfer term, $T_{SGS}=\tl{m}_{ij}\partial_{j}\tl{u}_{i}$, which represents the energy transfer from the resolved to subgrid scales \cite{Hartel1994}. Note that $\tl{m}_{ij} = H^{-1}\left( m_{ij} \right)$, where $\tl{m}_{ij}$ is defined by equation \eqref{sgsstress}. Plots of $T^{+}_{SGS}$ as a function of the wall normal distance are shown in Figure \ref{TSGSprofiles}. Both the total transfer due to the $\tl{m}_{ij}$ term and the individual contributions from the $\tl{A}_{ij}$, $\tl{B}_{ij}$ and $\tl{C}_{ij}$ terms (see equation \eqref{sgsstress}) are shown. Note that the contributions from both the $\tl{A}_{ij}$ and $\tl{B}_{ij}$ terms are net dissipative, while the $\tl{C}_{ij}$ term produces net backscatter. For all three terms the instantaneous values (not shown) fluctuated about these mean values by an order of magnitude, exhibiting both forward transfer and backscatter. 
\\
\\
In Figure \ref{TSGSprofiles} for the case where $C=1/6$ (filter width of twice the grid size) the minimum subgrid transfer (or maximum SGS dissipation) is $T_{SGS}^{+}\approx -0.06$ which is in good agreement with that reported in the literature from filtering DNS data for a channel flow at the same Reynolds number with the same filter \cite{Hartel1998,Piomelli1996}. Thus, instead of excessive backscatter or insufficient dissipation, the main problem instead is that the dominant physics is too close to the wall. The peak transfer in our simulations occurs at $y^{+}\approx 5$, as compared to that in the literature at $y^{+}\approx 10$.

\subsubsection{Model bias towards tilting voriticity in the near wall region}
\label{modelbias}
\begin{figure}
\psfrag{vort}{$\omega_{x}^{+}$}
\psfrag{yplus}{$y^{+}$}
\mbox{
\subfigure[Streamwise]{\includegraphics[width=10cm,height=6cm]{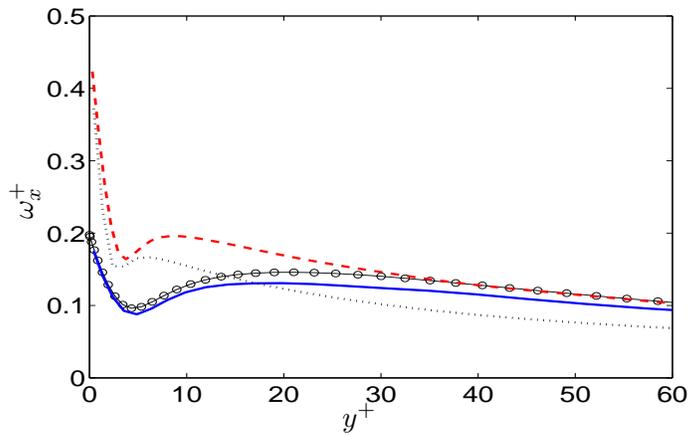}}}
\psfrag{vort}{$\omega_{z}^{+}$}
\mbox{
\subfigure[Spanwise]{\includegraphics[width=10cm,height=6cm]{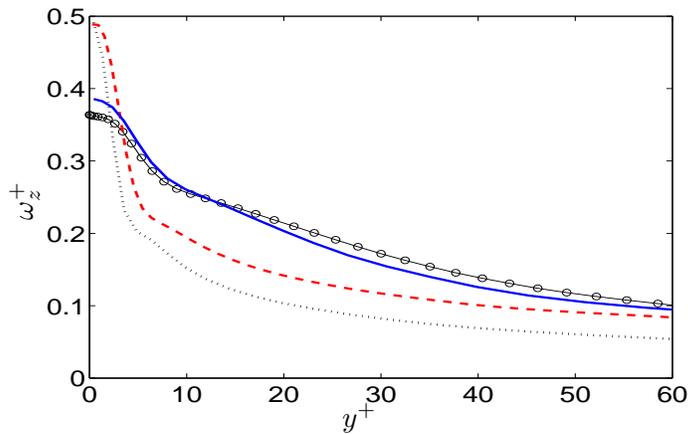}}}
\caption{(a) Streamwise and (b) Spanwise vorticity profiles. Solid line, no model $(24,64,24)$; dashed line, NS-$\alpha$ model $(32,128,32)$; dotted line, NS-$\alpha$ model $(24,64,24)$; symbols full channel DNS \cite{Kim1987}. The peak corresponding to the streamwise vortices in (a) is very close to the wall, $y^{+} \approx 9$, for the NS-$\alpha$ model. The spanwise vorticity is also overpredicted.}
\label{minvort}
\end{figure}

\begin{figure}
\psfrag{zplus}{$z^{+}$}
\psfrag{Ruuz}{$Ruu_{z}$}
\mbox{
\subfigure[DNS]{\includegraphics[width=8cm,height=6cm]{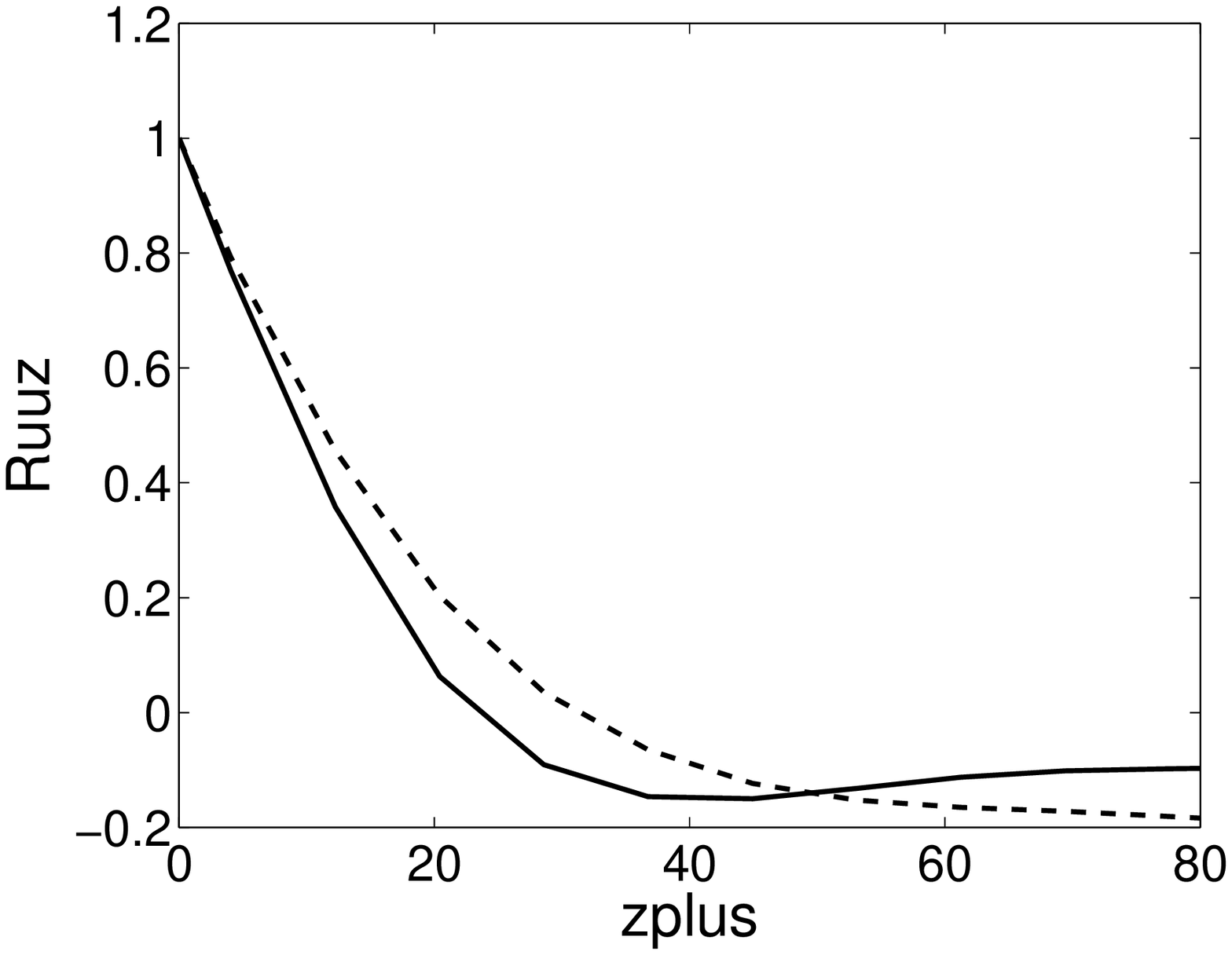}}}
\mbox{
\subfigure[NS-$\alpha$]{\includegraphics[width=8cm,height=6cm]{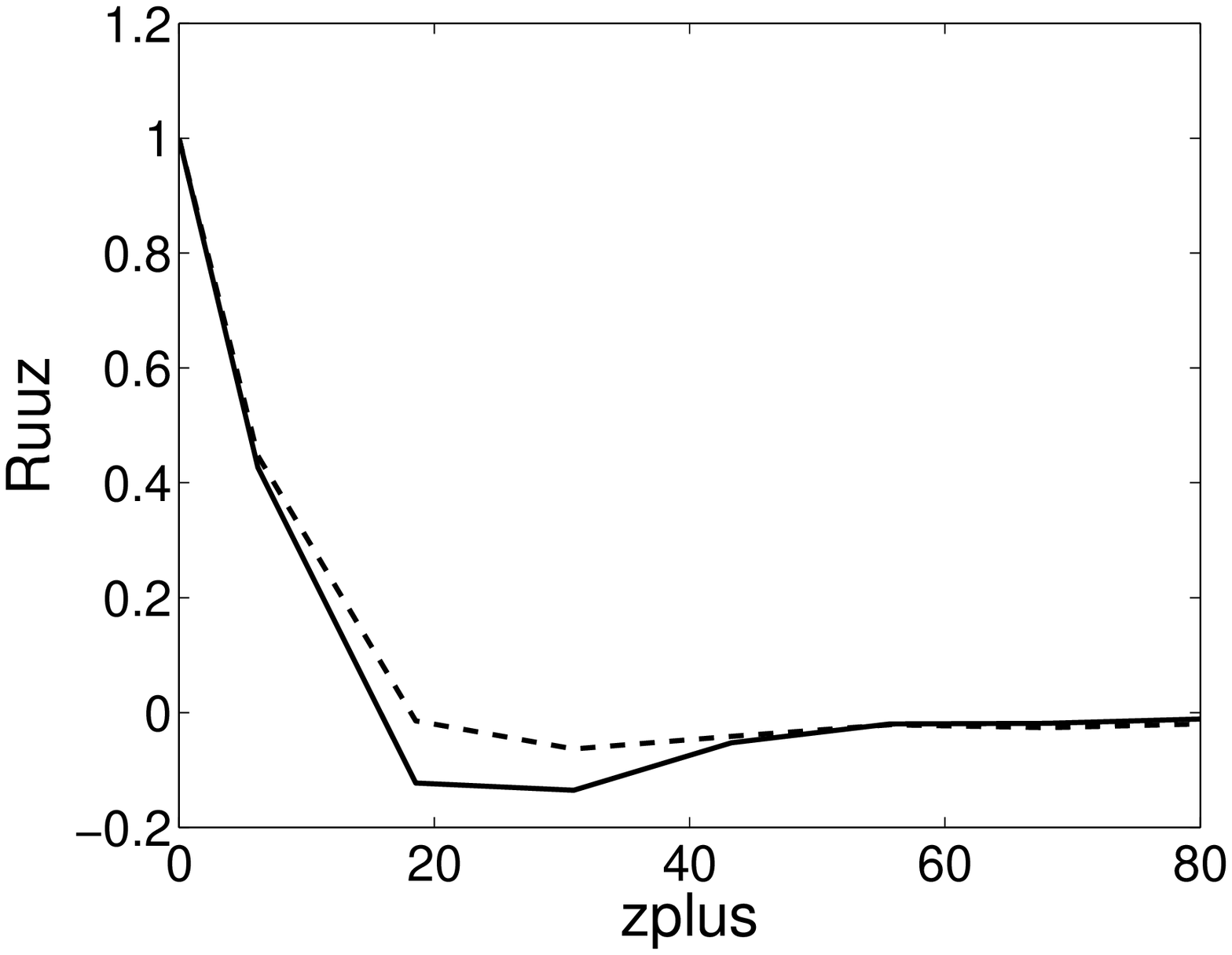}}}
\caption{Two point correlations of the streamwise velocity in the spanwise direction. Left panel is DNS and right panel is the NS-$\alpha$ model. Solid line is at $y^{+}\approx 7$, dashed line is $y^{+}\approx 18$. The streak spacing, indicated by the minimum of the two point correlation, is much smaller for the NS-$\alpha$ model than for the DNS.}
\label{twopointz}
\end{figure}

\begin{figure}
\psfrag{uplus}{$u^{+}$}
\psfrag{yplus}{$y^{+}$}
\psfrag{urms}{$u_{rms}$}
\includegraphics[width=12cm,height=8cm]{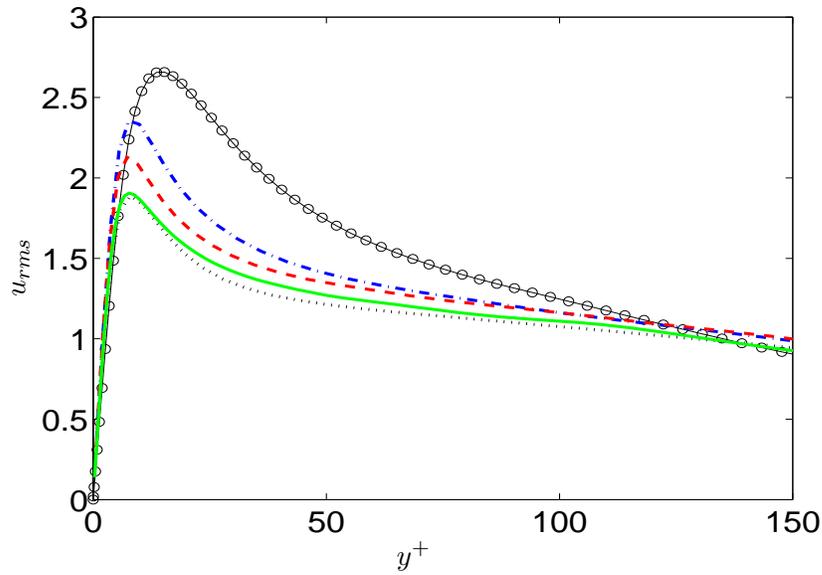}
\caption{Streamwise velocity fluctuations. Dash-dotted line, $C=1/6$, $(16,64,16)$; dashed line,  $C=1/6$, $(24,64,24)$; dotted line, $C=2/3$, $(32,64,32)$; solid line  $C=1/6$, $(32,128,32)$; symbols full channel DNS \cite{Kim1987}. To look at the effect of refining the mesh while keeping the subgrid resolution the dash-dotted, dashed and solid lines can be compared. To look at the effect of keeping the physical size of $\alpha^{2}_{k}$ constant while refining the mesh (hence increasing the subgrid resolution) the dash-dotted and dotted lines can be compared. }
\label{minrms}
\end{figure}

\begin{figure}
\psfrag{P(theta)}{$P(\theta)$}
\psfrag{theta}{$\theta$}
\mbox{
\subfigure[$y^{+}\approx 7.0$]{\includegraphics[width=8cm,height=6cm]{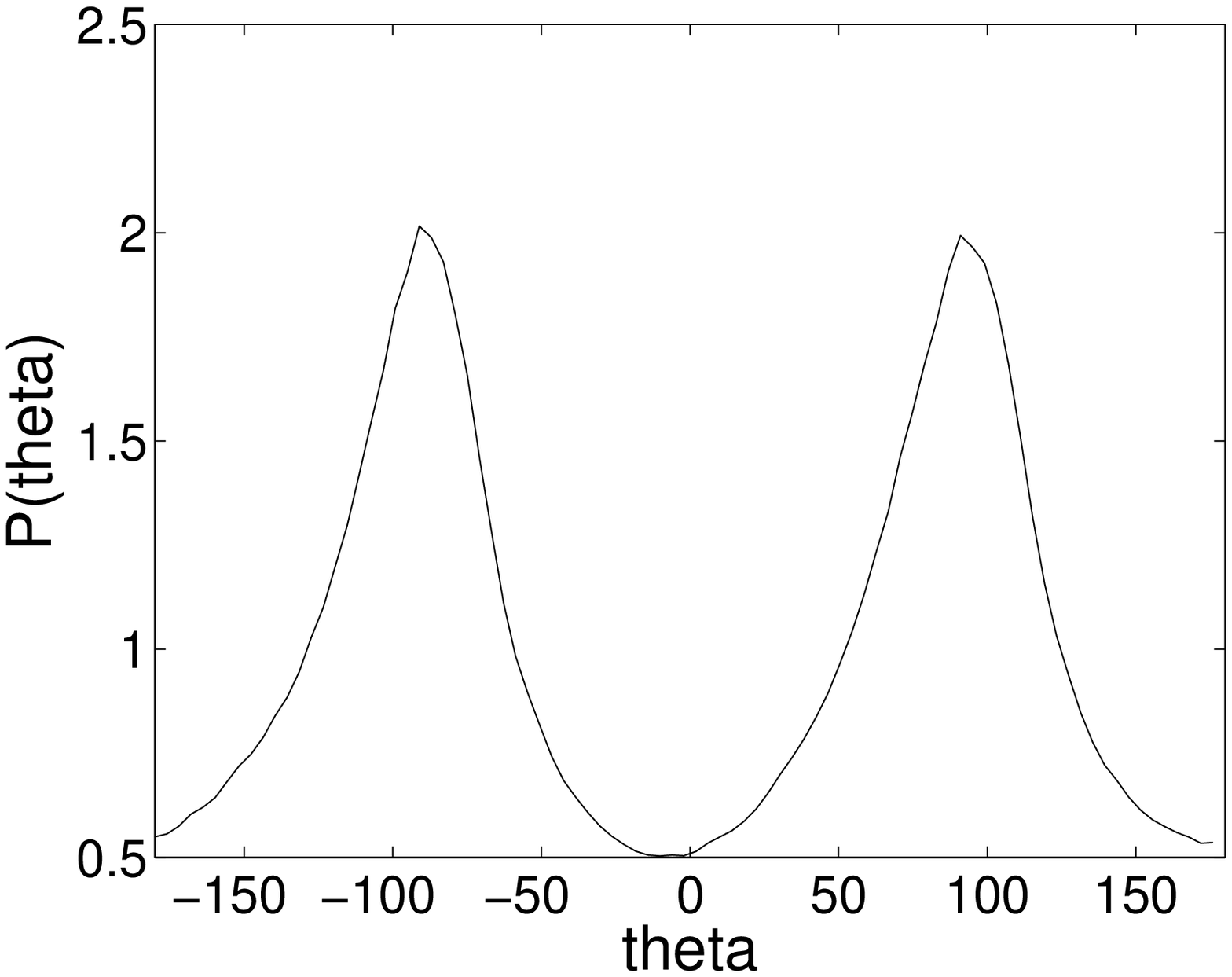}}}
\mbox{
\subfigure[$y^{+}\approx 5.4$]{\includegraphics[width=8cm,height=6cm]{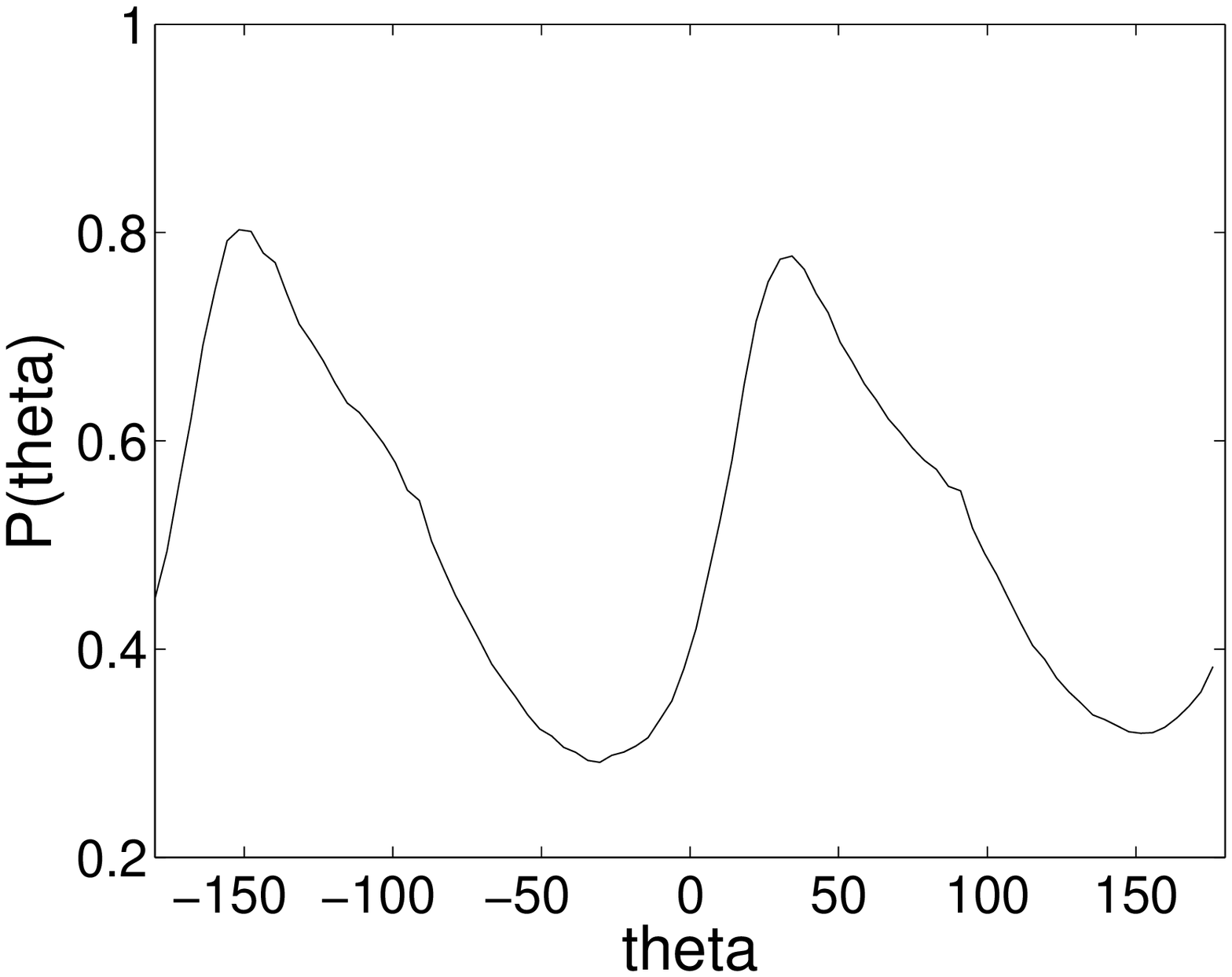}}}
\mbox{
\subfigure[$y^{+}\approx 18$]{\includegraphics[width=8cm,height=6cm]{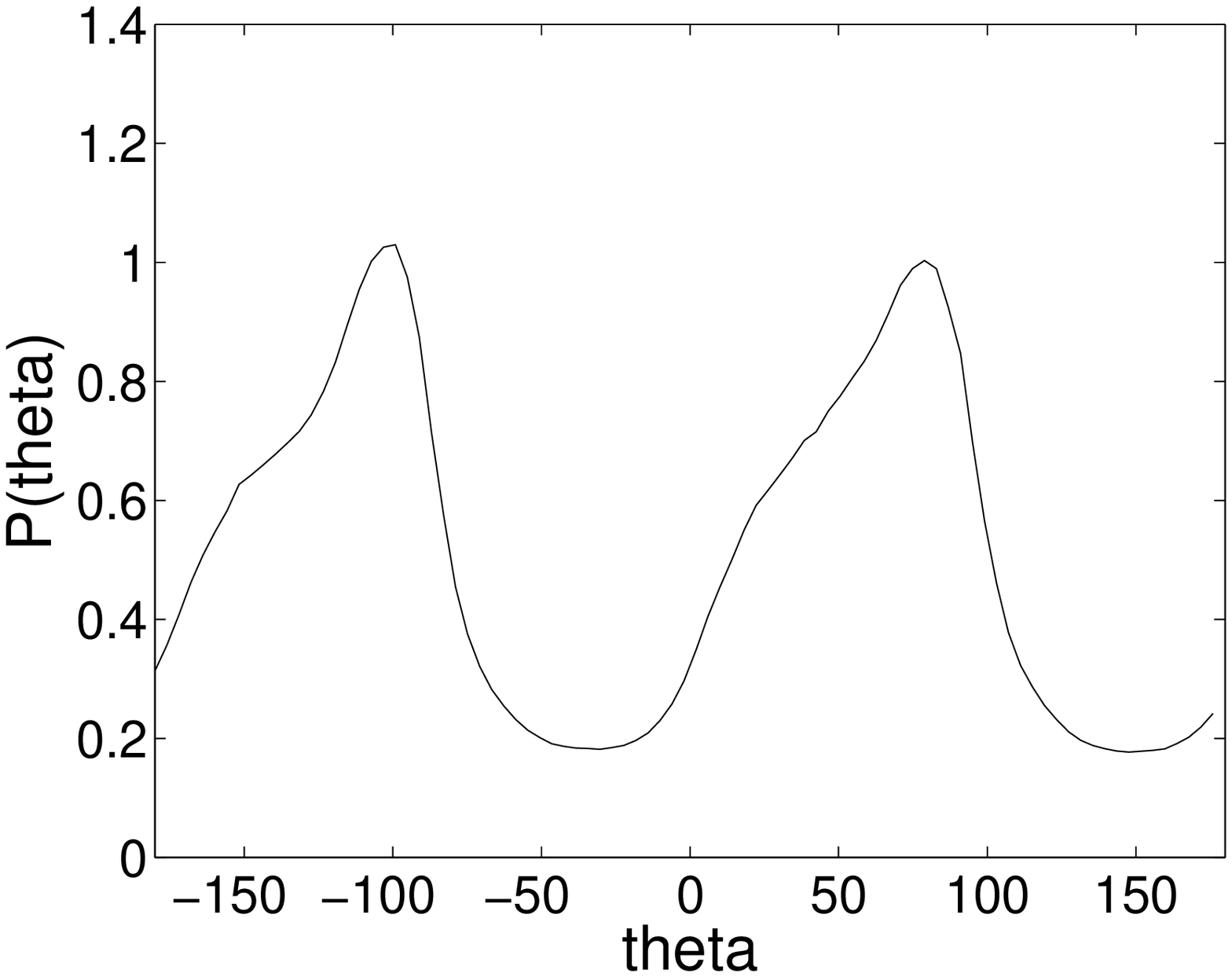}}}
\mbox{
\subfigure[$y^{+}\approx 9.7$]{\includegraphics[width=8cm,height=6cm]{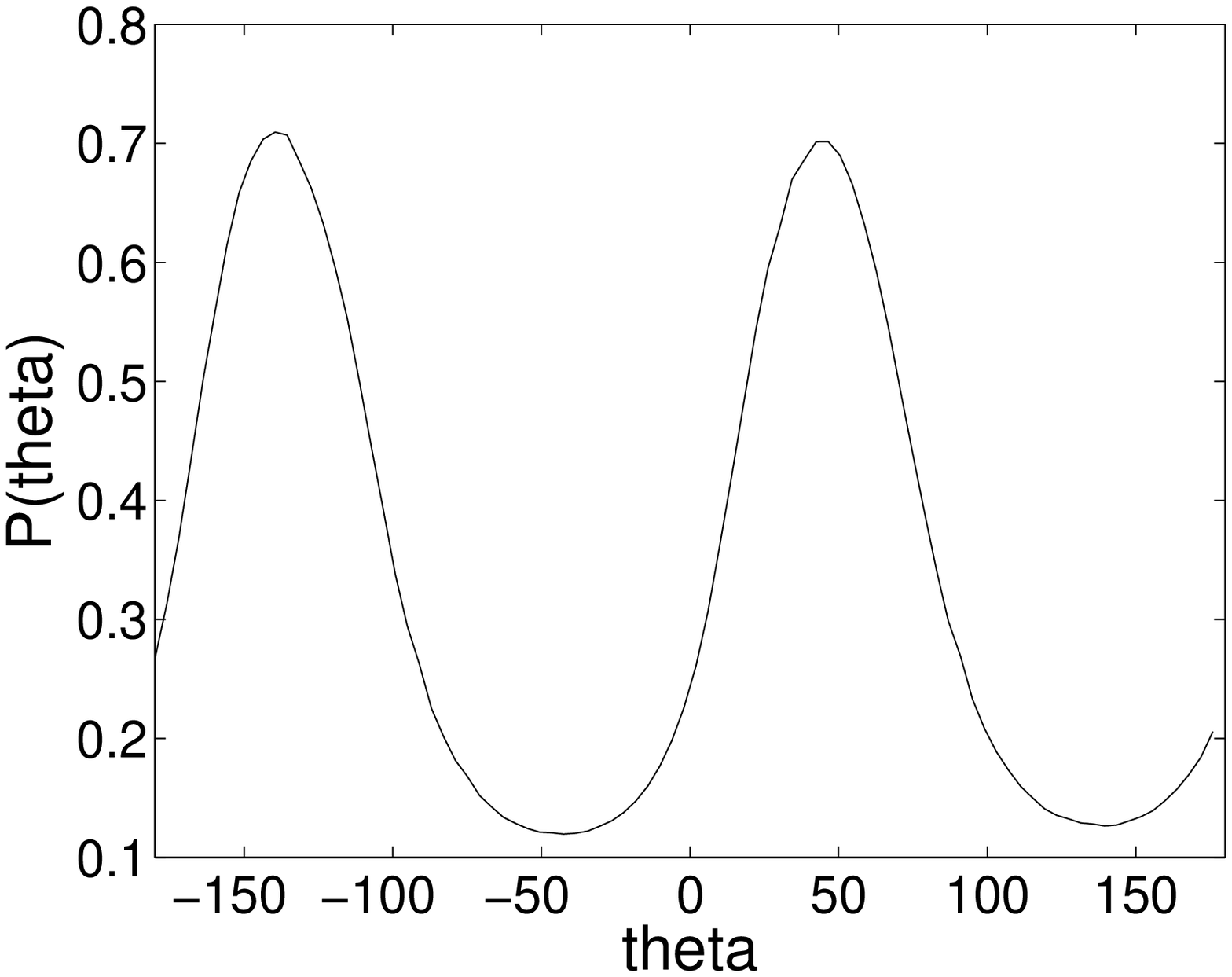}}}
\caption{Streamwise vorticity PDFs. Left column is DNS and right column is NS-$\alpha$ model. Notice that the peak corresponding to streamwise vortices (at $\approx 20^{o}$ is very prominent for the NS-$\alpha$ model close to the wall at $y^{+}=5.4$, while it doesn't appear in the DNS until $y^{+}=18$. }
\label{vortpdfs}
\end{figure}

\noindent
There is a strong correlation between the strength of the streamwise vortices and skin friction \cite{Kravchenko1993}. We show here that the vorticity field produced by the NS-$\alpha$ model is erroneous, and that this is the cause of the high skin friction (which manifests itself as an underpredicted mean flow profile, shown in Figure \ref{minulog}).
\\
\\
Streamwise and spanwise vorticity fluctuations,  $\omega^{+}=\omega_{rms} \nu/u_{\tau}^{2}$, are shown in Figure \ref{minvort} for the $(24,64,24)$ and $(32,128,32)$ meshes, both use $C=1/6$.  The minimal channel DNS is in good agreement with the data from the full channel DNS \cite{Kim1987}, while the NS-$\alpha$ model significantly overpredicts the streamwise and spanwise vorticity fluctuations very close to the wall. The peak in the streamwise vorticity fluctuation at the edge of the buffer layer (at $y^{+} \approx 20$ for the DNS) is much closer to the wall for the NS-$\alpha$ model (here at $y^{+} \approx 9$) and higher in magnitude. This peak is indicative of the streamwise vortices in the buffer layer \cite{Kim1987}. According to the streamwise vortex model of Kim et al. \cite{Kim1987}, the ratio between the streamwise vorticity peak at the wall to that in the buffer layer should be $\approx 1.3$. Here we have instead  a ratio of $\approx 2$ (taking the peak at $y^{+} \approx 10$ to be the buffer layer vortices). The wall value of the streamwise vorticity  for the NS-$\alpha$ model is $\omega^{+}_{x}|_{w}\approx 0.4$, twice that in the DNS where $\omega^{+}_{x}|_{w} \approx 0.18$. 
\\
\\
In the Introduction it was highlighted that in the NS-$\alpha$ model vortices are tilted and stretched by a smoothed velocity, (see equation \eqref{NSalphavort}). In the near wall region streamwise vorticity is primarily created by tilting of the spanwise vorticity into the streamwise direction, through the $\omega_{z}\frac{\partial \tl{u}}{\partial z}$ term in the spanwise vorticity equation. Writing the smoothed velocity as $\tl{u}_{i}=u_{i}+\alpha^{2}_{k} \nabla^{2}\tl{u}_{i}$ we can see that the term responsible for tilting spanwise vorticity into the streamwise direction, $\omega_{z}\frac{\partial u}{\partial z}$, is augmented by $\omega_{z} \partial_{z} \left( \alpha^{2}_{z}\partial^{2}_{z}\tl{u}\right)$. We can compare the NS-$\alpha$ tilting term with that from the Navier-Stokes equation by writing the tilting term for the Navier-Stokes equation as
\begin{equation}
\omega_{tiltx}^{NS} \approx \omega_{z}^{NS}\left(\frac{\partial u}{\partial z}\right)^{NS},
\end{equation}
and that for the NS-$\alpha$ equation as
\begin{equation}
\omega_{tiltx}^{\alpha} \approx \omega_{z}^{\alpha}\left(\frac{\partial \tl{u}}{\partial z}\right)
                                                 +\omega_{z}^{\alpha}\frac{\partial }{\partial z}\left(\alpha^{2}_{z}\frac{\partial^{2} \tl{u}}{\partial z^{2}}\right).
\end{equation}                                                 
The most significant contribution to the spanwise velocity gradient, $\partial u/\partial z$, close to the wall is from the streaks, thus we take this velocity gradient to be proportional to the rms streamwise velocity fluctuation divided by the streak spacing. The streak spacing can be measured from the two point correlation $R_{uuz}$ \cite{Kim1987}, shown here in Figure \ref{twopointz}. We found the streak spacing normalized by $(\nu/u_{\tau})$ to be narrower with the NS-$\alpha$ model, at  $\approx40$, than the DNS result by approximately a factor of two
\footnote{It is not clear at this point why this is the case. One possibility is, given that the streak spacing is believed to emerge from a secondary instability of the Tollmein-Schlicting wave (Jimenez pg. 219 \cite{Jimenez1991}), the spacing we see here may be related to possible differences that would arise through a stability analysis of the NS-$\alpha$ equation as compared to the same anaylysis for the Navier-Stokes equation.  For example, it has been shown that the model lowers the critical wavenumber for baroclinic instability in a two-layer quasi-geostrophic model. Although the initialization here was not representative of a true transition process, there were significant differences observed in how the flow became turbulent from the perturbed laminar state when the NS-$\alpha$ model was used, as compared to without.}.
\\
\\
Using the streamwise velocity values from Figure \ref{minrms}, for the minimal channel DNS $\partial u/\partial z$ is approximately $2.7/80=0.035$, while for the NS-$\alpha$ model it is $2/40=0.05$. If we then take the contribution for the second term, $\partial_{z} \left(\alpha^{2}_{z}\partial^{2}_{zz}\tl{u}\right)$, to be proportional to $(\alpha_{z}^{2}/h_{z}^{2})(d \tl{u}/dz)$ we arrive at the following relationship between the two source terms
\begin{equation}
\frac{\omega_{tiltx}^{\alpha}}{\omega_{tiltx}^{NS}}=
\frac{\omega_{z}^{\alpha}(d\tl{u}/dz)(1+\alpha^{2}_{z}/h_{z}^{2})}
        {\omega_{z}^{NS}(du/dz)^{NS}}.
\label{tiltratio}
\end{equation}
Substituting $(d\tl{u}/dz)/(du/dz)^{NS} \sim 0.05/0.035$, $\alpha^{2}_{z}/h_{z}^{2}\approx 1/6$ and $\omega_{z}^{\alpha}/\omega_{z}^{NS}=0.48/0.38$ (wall values from Figure  \ref{minvort}) into Equation \eqref{tiltratio} we arrive at
\begin{equation}
\frac{\omega_{tiltx}^{NS}}{\omega_{tiltx}^{\alpha}}=2.1
\end{equation}
which agrees well with the values from the minimal channel of $0.40/0.18$. 
\\
\\
To investigate the streamwise vortices in the NS-$\alpha$ model further, in Figure \ref{vortpdfs} we compare probability density functions (PDFs) of the streamwise vortex inclination angle $\theta=\arctan(\tl{\omega}_{y}/\tl{\omega}_{x})$ at two different heights from the wall. For the DNS the PDFs were measured using the instantaneous vorticity vector on two $x-z$ planes, at vertical locations of $y^{+}\approx 7$, and $y^{+}\approx 18$. The first is in the viscous sublayer,while  the second is in the buffer layer. In the viscous sublayer we can see two peaks in the PDF at $\pm 90^{o}$  corresponding to the low and high speed streaks.  As you move into the buffer region a shoulder appears near $25^{o}/-155^{o}$ that corresponds to the streamwise vortices. These results are in good agreement with those from the literature \cite{Moin1985}.
\\
\\
For the NS-$\alpha$ model  to see evidence of streamwise vortices, the PDFs needed to be measured closer to the wall, and are shown in Figure \ref{vortpdfs}. At $y^{+} \approx 5.4$ we can see there are shoulders near $\theta \approx \pm 90^{o}$ and more distinct peaks at $\theta \approx +20^{o}/-160^{o}$, the former indicating streamwise streaks and the latter indicating streamwise vortices. PDFs measured closer to the wall (not shown) had a single peak at $\theta \approx 0$ (zero vertical vorticity, as required by the no-slip condition at the wall). Thus we did not see any PDFs indicating a region dominated by low-speed streaks (similar to the one at $y^{+} \approx 7$ for the DNS), but instead the region close to the wall shows a dominant signature of streamwise vortices. 
%Vortices that are more inclined may lead to higher skin friction. See PoF paper by Lee and Kim on control of the viscous sublayer to reduce drag vol 14, 2523, 2002. 

\subsubsection{Using damping to overcome the model bias}
\label{biasresp}
\begin{figure}[tbp]
\psfrag{uplus}{$u^{+}$}
\psfrag{yplus}{$y^{+}$}
\psfrag{urms}{$u_{rms}$}
\psfrag{vrms}{$v_{rms}$}
\psfrag{wrms}{$w_{rms}$}
\psfrag{uv}{$\ov{u'v'}$}
\mbox{
\subfigure[Mean flow]{\includegraphics[width=8cm,height=6cm]{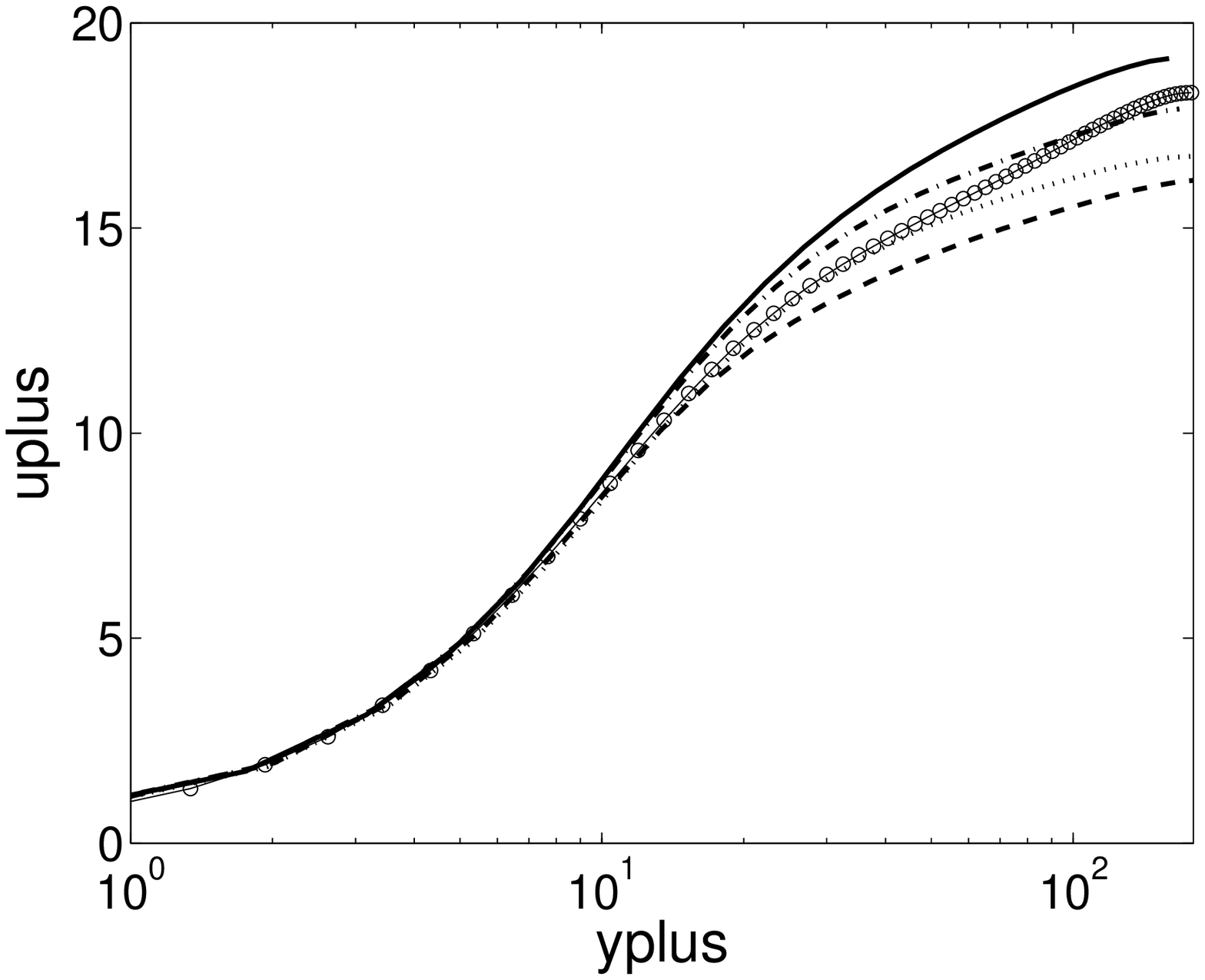}}}
\mbox{
\subfigure[Streamwise]{\includegraphics[width=8cm,height=6cm]{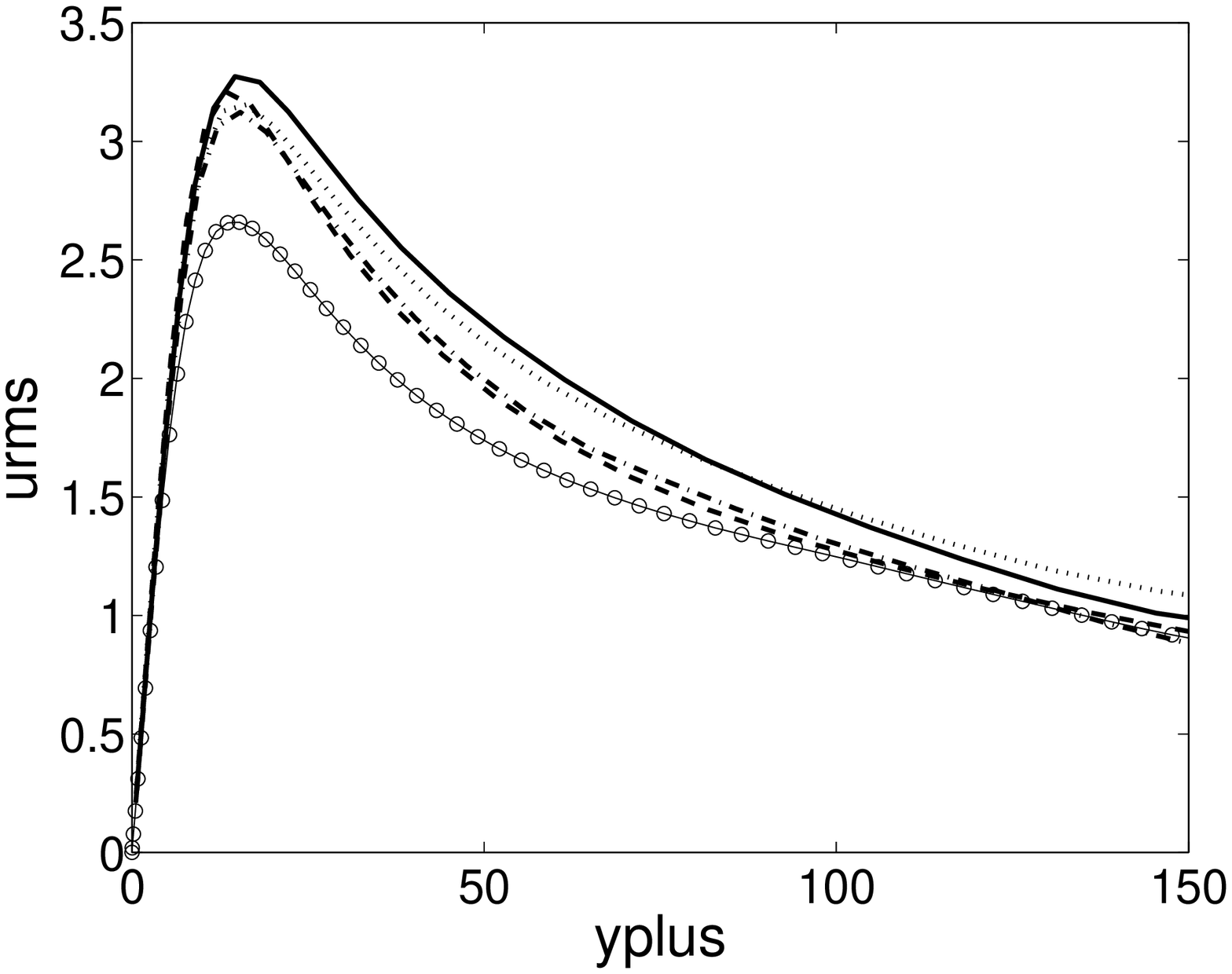}}}
\mbox{
\subfigure[Vertical]{\includegraphics[width=8cm,height=6cm]{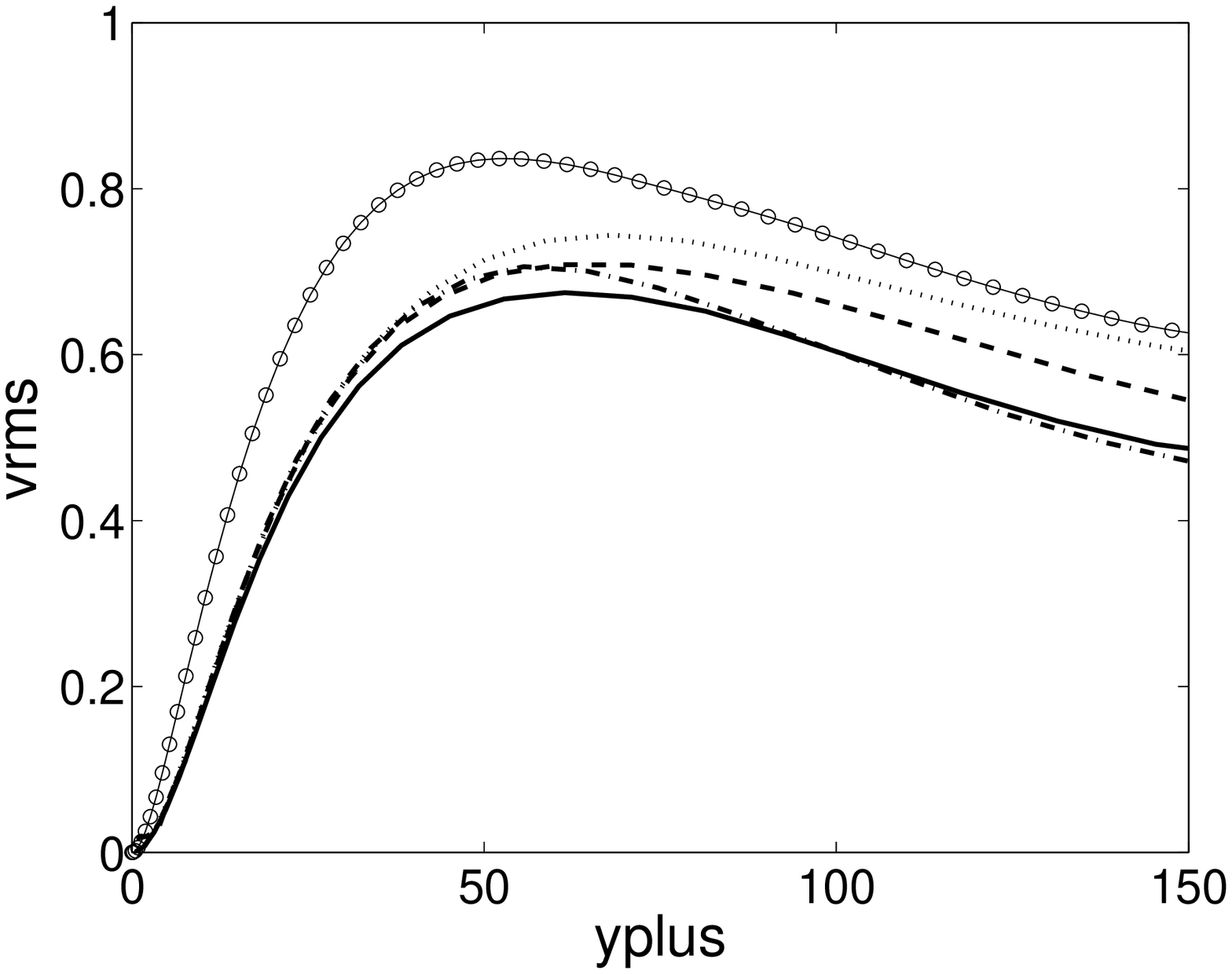}}}
\mbox{
\subfigure[Spanwise]{\includegraphics[width=8cm,height=6cm]{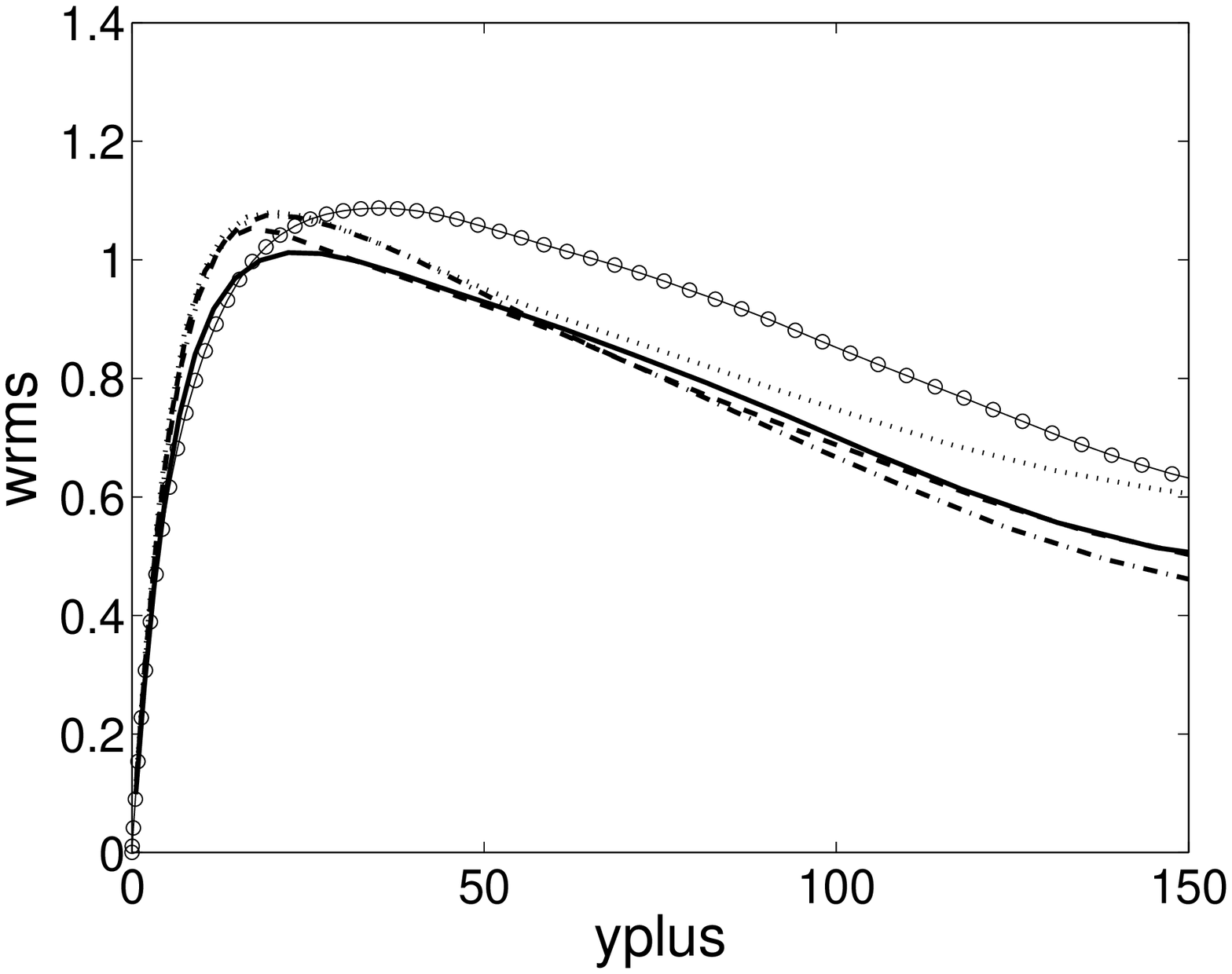}}}
\mbox{
\subfigure[Shear stress]{\includegraphics[width=8cm,height=6cm]{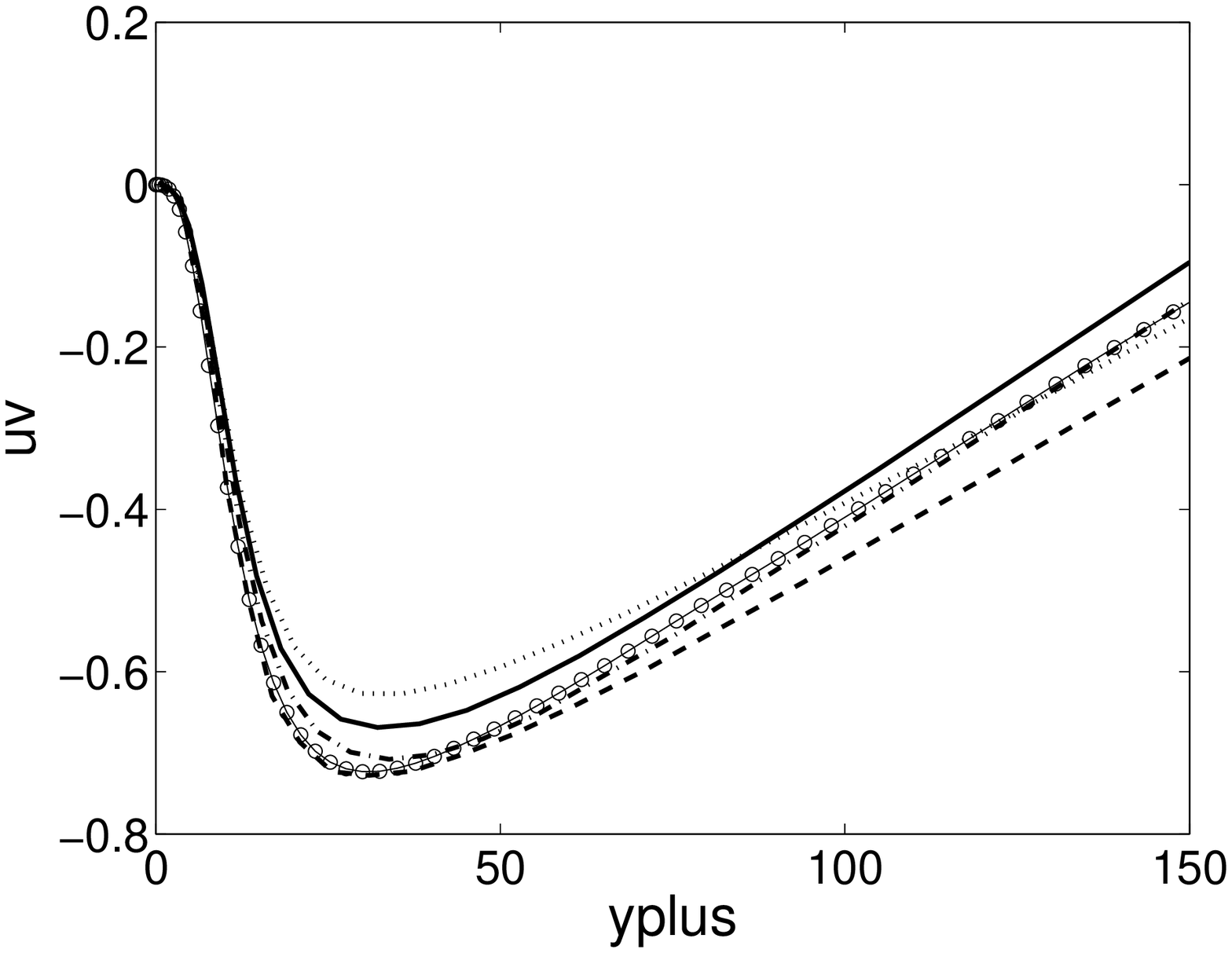}}}
\caption{Mean flow, rms and shear stress profiles; no model (solid); anisotropic NS-$\alpha$ with damping (dashed); isotropic NS-$\alpha$ with damping (dash-dot); anisotropic Leray without damping (dotted). Symbols are DNS data \cite{Kim1987}. Note in particular that the mean flow profile and skin friction for the damped isotropic model are in very good agreement with the DNS.}
\label{rmsdamp}
\end{figure}

\noindent
In the previous section we found that the NS-$\alpha$ model provides an additional mechanism for producing atreamwise vorticity in the near-wall region by tilting spanwise vorticity directly into the streamwise direction. This is physically  incorrect in two respects. First, vortex tilting and stretching processes should occur farther away from the wall, in the buffer region, not in the viscous sublayer which is what we see here. Second, the path of streamwise vorticity creation is incorrect.  In the literature  the streamwise vorticity comes from \textit{first} lifting up the transverse vorticity into the buffer layer and \textit{then} tilting of the vorticity into the streamwise direction \cite{Jimenez1991}, while here we have a direct tilting of spanwise vorticity into the streamwise direction. This indicates that either damping of $\alpha^{2}_{k}$ in the near wall region or an alternative specification of $\alpha^{2}_{k}$ is necessary. Here we investigate damping, alternative definitions of $\alpha^{2}_{k}$ are left to a future study.
\\
\\
Zhao and Mohseni \cite{Zhao2005a} found in an \textit{a priori} study using a dynamic procedure that $\alpha$ followed a linear variation from zero at the wall to a constant value of $0.02$ around $y^{+} \approx 10$.  In a later study \cite{Zhao2005b} they tested this distribution for $\alpha$ (now turning off the dynamic procedure and fixing $\alpha$ to follow the specified profile), but their results showed a significant overprediction of the spanwise velocity fluctuations and  also some overprediction of the vertical velocity fluctuations (together suggesting high streamwise vorticity). Otherwise their results were not significantly better than a standard LES.  This is not surprising because their $\alpha$ distribution was determined using an \textit{a priori} study, which does not account for the feedback of the model on the flow. For example, in the previous section we saw that the effect of having the streamwise vortices closer to the wall is to increase skin friction. This is the type of effect you will not see in an \textit{a priori} study. 
\\
\\
Following Zhao and Mohseni we used a linear variation of $\alpha$ and $\alpha_{k}$ from zero at the wall to a constant value at a specified wall-normal distance. Here the damping was applied over the region $y^{+} < 60$ instead of the region $y^{+} <10$ that was used in their study. This choice was motivated by the study by Jimenez that demonstrated in the region $y^{+} < 60$ the low-speed streaks are a critical part of the autonomous cycle of near-wall turbulence \cite{Jimenez1999b}. Thus we consider this to be the `streak-affected' region, and since the problem is related to the velocity gradients from the streaks, $\partial \tl{u}/\partial z$, it is logical to apply the damping factor through this region. Other values were tested along with exponential damping profiles instead of the simple linear one. The shape of the damping profile was found to be insignificant, with the wall-normal distance being the important factor. Damping over the streak affected region was found to consistently provide the best results. The damping function used was
\begin{equation}
f(y^{+})=\left\{\begin{array}
{r@{\quad \quad}l}
\left(y^{+}/60\right)^{2} & \mbox{if} \quad y^{+} \le 60 \\
1 & \mbox{otherwise.}
\end{array}\right. 
\label{dampzhao}
\end{equation}
Results with damping  are shown in Figure \ref{rmsdamp} for the isotropic and anisotropic models. These are now for full channels with $(L_{x}, L_{y},L_{z})=(4\pi,2,2\pi/3)$ and $(N_{x},N_{y},N_{z})=(32,48,32)$. For the isotropic model $\alpha^{2}$ was specified as $\alpha^{2}=f\left(y^{+}\right)(0.02)^{2}$, where $f\left(y^{+}\right)$ is given by equation \eqref{dampzhao} and $0.02$ is the value of $\alpha$ away from the wall determined by Zhao and Mohseni \cite{Zhao2005a}. We also tried a value of 0.04, which they determined for the $Re_{\tau}$=180 channel from scaling arguments, but we found this was too high to yield reasonable results. 
\\
\\
For the anisotropic model $\alpha^{2}_{k}=f \left(y^{+}\right)C h_{k}^{2}$ was used for $\alpha^{2}_{x}$ and $\alpha^{2}_{z}$ while $\alpha^{2}_{y}$ was set to zero. Initially a $C$ value of $1/6$ was used such that it is a damped version of the case where the filter width is twice the grid spacing. However, with this value the logarithmic law was still underpredicted (in terms of the $y$-intercept), so the results reported here used $C=1/12$. Even with this value the skin friction is still overpredicted. To have the same physical equivalent $\alpha$ for the isotropic and anisotropic models you would need to use $C=1/24$ or  $\Delta=h$. This was tested (results not shown) and it did bring the mean velocity profile into good agreement with the DNS data, but it does not seem to make good physical sense because this would mean the filter width is equal to the grid spacing. 
\\
\\
We can see in Figure \ref{rmsdamp} that damping removes the problem with the high spanwise fluctuations, and improves some quantities slightly (eg. mean flow profile, shear stress and streamwise velocity fluctuations) but overall the differences between the no model and NS-$\alpha$ model results are very small when damping is used. In Figure \ref{rmsdamp} we also show results from a simulation using the anisotropic Leray model (with $\partial_{j}\tl{u}_{j}=0$ enforced), that is the $\tl{A}_{ij}+\tl{B}_{ij}$ terms in equation \eqref{sgsstress}. The Leray model does not have the same vortex tilting properties that the NS-$\alpha$ model has. We can see by comparing the Leray model results in Figure \ref{rmsdamp} with those from the NS-$\alpha$ model in Figures \ref{minulog} and \ref{minrms} that the Leray model does not suffer the same underprediction of the mean flow that the NS-$\alpha$ model does, reinforcing the fact that it is the $C_{ij}$ term that is causing the problem. The tilting term, $u_{k}\partial_{i}\tl{u}_{k}$, which combines with the modified pressure gradient in equation \eqref{NSalpha} to form the $\partial_{j}(C_{ij})$ term in the model, is the unique feature of the NS-$\alpha$ model. 

\subsubsection{Helicity PDFs}
\begin{figure}[tp]
\psfrag{P(h)}{$P(\tl{h})$}
\mbox{
\subfigure[minimal channel, $y^{+}\approx 90$]{
\includegraphics[height=6cm]{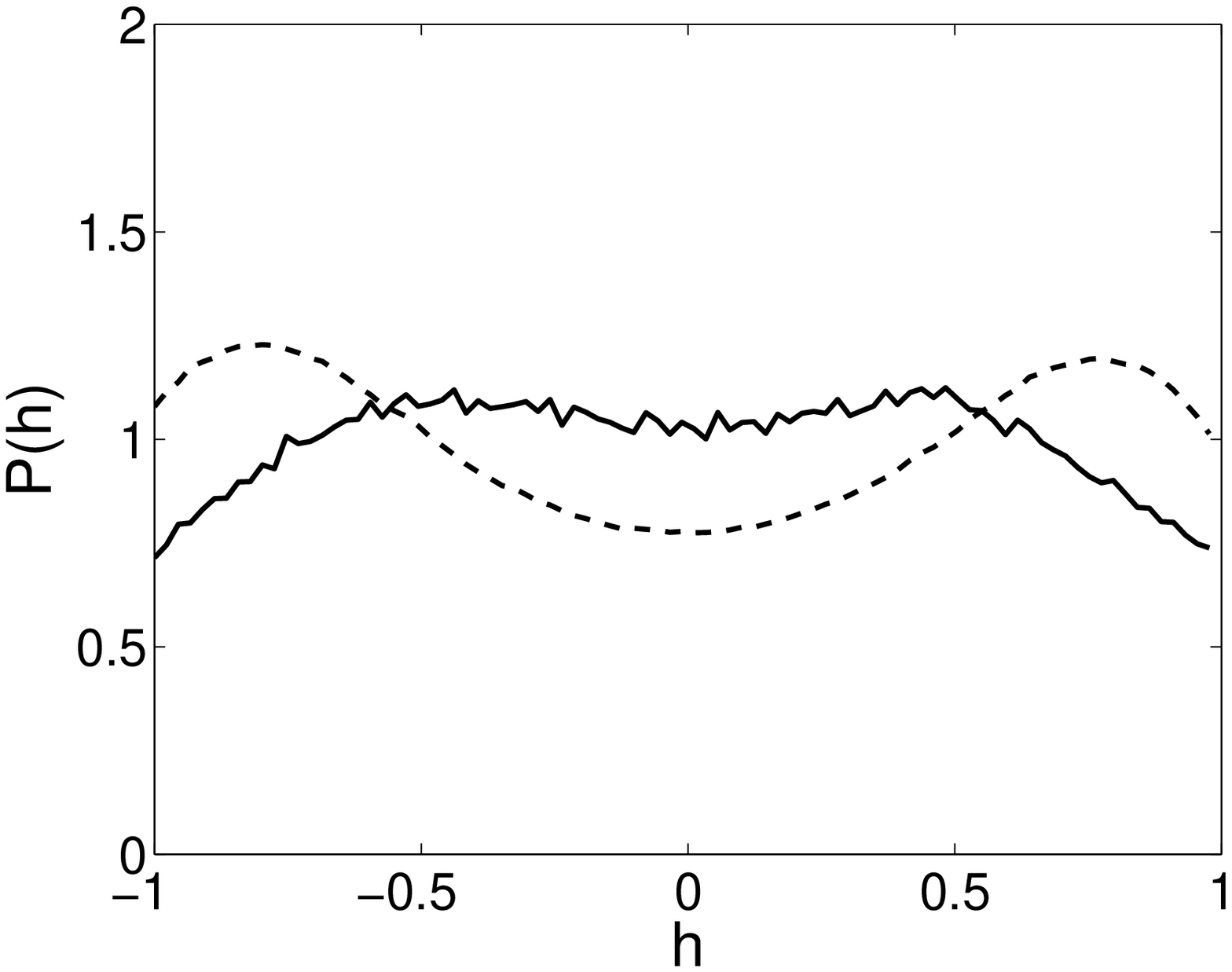}}}
\mbox{
\subfigure[full channel, $y^{+}\approx 90$]{
\includegraphics[height=6cm]{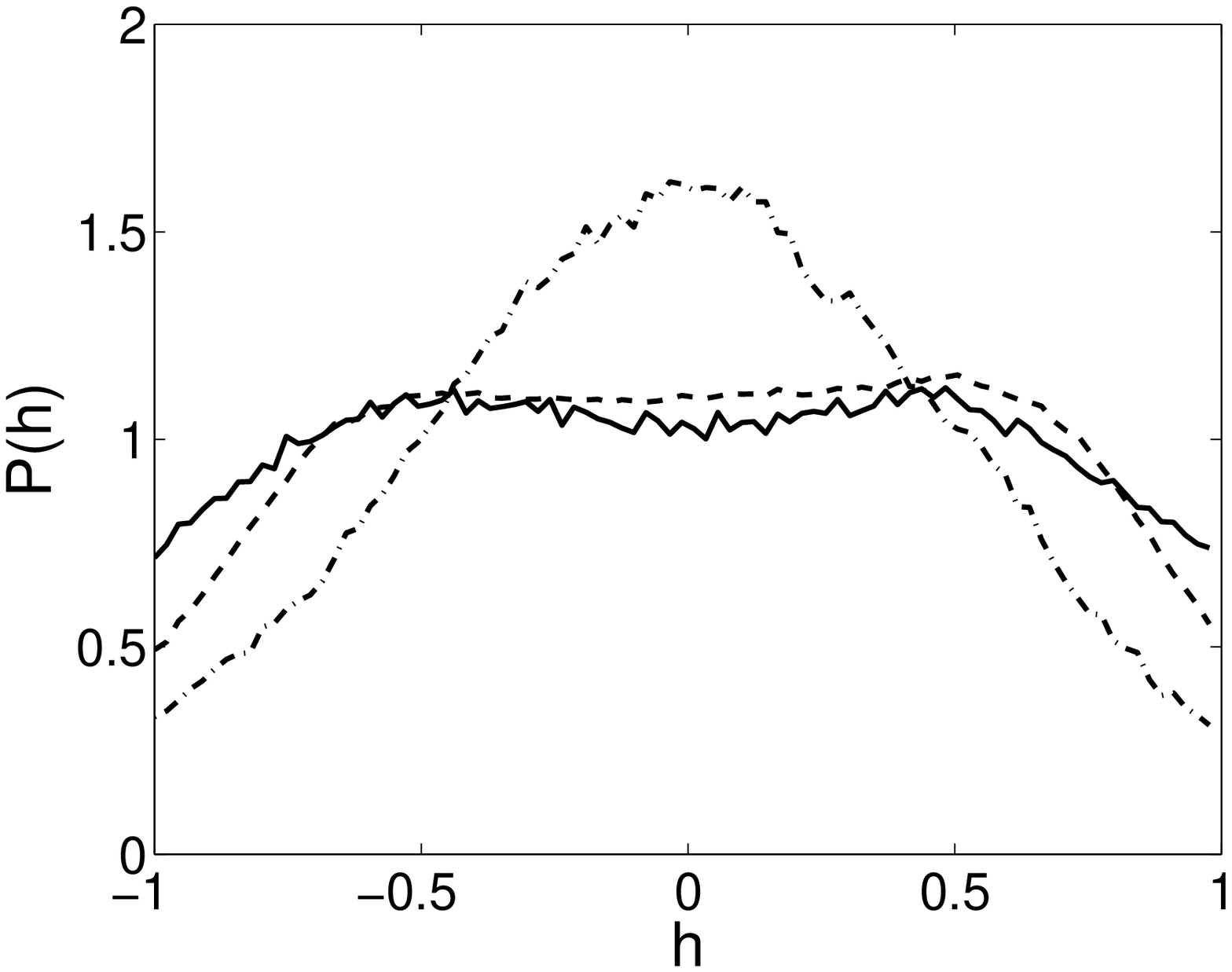}}}
\caption{PDFs of the relative helicity, (a) minimal channel; DNS (mesh IV), solid; NS-$\alpha$ (mesh IV), dashed; (b) full channel; NS-$\alpha$ with damping (anisotropic model), dashed;  no model, dash-dotted;. solid line is from the minimal channel DNS, shown for comparison. In (b) we can see that the NS-$\alpha$ helicity PDF is very similar to that from the DNS, whereas when no model is used the shape of the helicity PDF is incorrect.}
\label{helpdfs}
\end{figure}

\noindent
In the previous two sections we have seen that the NS-$\alpha$ model has a tendency to tilt vorticity close to the wall, and that with damping this impact is reduced. We now address the question of how the model changes the vorticity and velocity fields away from the wall. To do this we look at the relative helicity, defined as  \cite{Rogers1987}
\begin{equation}
h = \frac{\boldsymbol{u} \cdot \boldsymbol{\omega}}{| \boldsymbol{u} | |\boldsymbol{\omega}|}.
\end{equation}
The helicity is related to the nonlinear term $\bs{u}\times\bs{\omega}$ through the identity\begin{equation}
\frac{\left(\bs{u}\cdot \bs{\omega}\right)^{2}}
{|\bs{u}|^{2}|\bs{\omega}|^{2}}+
\frac{\left(\bs{u}\times \bs{\omega}\right)^{2}}
{|\bs{u}|^{2}|\bs{\omega}|^{2}}=1.
\end{equation}
Thus by looking at the helicity, we can also examine the non-linearity of the model. Given that the NS-$\alpha$ equations are described as having a reduced nonlinearity \cite{Domaradzki2001},  we expect they may also have high helicity. For the NS-$\alpha$ model in the LES-template we can write
\begin{equation}
\frac{\partial \bs{\tl{u}}}{\partial t}+\bs{\tl{\omega}}\times\bs{\tl{u}}=-\nabla\left(p+\frac{1}{2}\bs{\tl{u}}\cdot\bs{\tl{u}}\right)+\nu \nabla^{2}\bs{\tl{u}}-H^{-1}\left(\frac{\partial m_{ij}}{\partial x_{j}}\right)
\end{equation}
and investigate the smoothed relative helicity
\begin{equation}
\tl{h} = \frac{\boldsymbol{\tl{u}} \cdot \boldsymbol{\tl{\omega}}}{| \tl{\boldsymbol{u}} | |\tl{\boldsymbol{\omega}}|}.
\end{equation}
The focus now is on the region away from the wall, thus the PDFs shown in Figure \ref{helpdfs} were measured  at $y^{+}\approx 90$. We can see for the minimal channel the NS-$\alpha$ model has two shoulders near $\pm 1$ indicating a higher probability of increased helicity (reduced nonlinearity) relative to the Navier-Stokes equations. Values of the mean-squared helicity, $\ov{\tl{h}^{2}}$, are given in Table \ref{heltable}. A uniform distribution would correspond to $\ov{\tl{h}^{2}}=0.33$. For the full channel we can see when no model is used the PDF is too peaked and $\ov{\tl{h}^{2}}$ is too low, but when the damped NS-$\alpha$ model is used the results are closer to the minimal channel DNS (which is itself close to the full channel DNS of Rogers and Moin \cite{Rogers1987}). Because we expect the minimal channel DNS to be representative of a full channel at a finer mesh spacing, this suggests that the NS-$\alpha$ model can produce helicity statistics on a coarse mesh that are comparable to those from a finer mesh without a model.
 
 \section{Conclusions}

In this paper the NS-$\alpha$ model has been investigated for a fully turbulent channel flow. To begin we derived the model using Hamilton's principle. Using this derivation it is straightforward to see how the model can be extended to different physical situations. For example, compressible flow or geophysical flows, how the model could be altered by using a higher order expansion in the definition of the velocity fluctuation, or different definitions of $\xi$. The definition of $\alpha_{k}^{2}$ used in practice should be consistent with that used in the derivation. For example, we found that in our application of the NS-$\alpha$ model to the channel flow, when $\alpha^{2}_{k}$ is based on the mesh the $\alpha^{2}_{z}$ values are too large and this leads to excessive tilting of spanwise vorticity into the streamwise direction in the near wall region. This is because we derived the model assuming that $\alpha_{k}^{2}$ follows an advection equation, and technically we should have solved an advection equation to determine $\alpha^{2}_{k}$, instead of basing it on the computational mesh. The fact that the magnitudes of the values used for $\alpha^{2}$ with damping are close the those one would obtain from solving an advection equation \cite{Armenio1999} reinforces this statement.
\\
\\
Given the significant impact the NS-$\alpha$ model has on the vorticity field, we feel that future studies should investigate how the NS-$\alpha$ model affects the resolved flow vortices (see da Silva et al. \cite{Silva2004} for a study of this nature for other subgrid models).  Such a study would have practical implications as well, for example in applications where it is the size, strength and location of the fluid vortices that is of interest.

\section*{Acknowledgements}
This work has been supported by the Natural Science and Engineering Research Council of Canada (NSERC) and Mathematics of Information Technology and Complex Systems (MITACS), and was made possible by the facilities of the Shared Hierarchical Academic Research Computing Network (SHARCNET) and Western Canada Research Grid (WESTGRID). We would also like to thank Kevin Lamb for his useful suggestions regarding this work, and the reviewers whose valuable comments greatly improved the organization of the manuscript.

\clearpage
\begin{table}[tbp]
\tbl{Mesh parameters for the minimal channel flow. The first column are the $(x,y,z)$ dimensions non-dimensionalized by the channel half-height, the second column is the number of mesh points in each direction, and the last three are the channel dimensions in wall units (non-dimensionalized by the shear velocity, $u_{\tau}$, and the viscosity, $\nu$).}
{\begin{tabular}{@{}lccccccc}\toprule
  & $(L_{x},L_{y},L_{z})$ & $(N_{x},N_{y},N_{z})$ & $h_{x}^{+}$ & $h_{y}^{+}(min)/h_{y}^{+}(max)$ & $h_{z}^{+}$ \\
\colrule
I  &    $(\pi,2,0.3\pi)$ &  (16,64,16)   & 35.3 & 0.875/11.6 & 10.6 \\
II &    $(\pi,2,0.3\pi)$ &  (24,64,24)   & 23.6 & 0.875/11.6 & 7.07 \\
III &    $(\pi,2,0.3\pi)$ &  (32,64,32)   & 17.7 & 0.875/11.6 & 5.30 \\
IV &   $(\pi,2,0.3\pi)$ & (32,128,32) & 17.7 & 0.424/5.80 & 5.30 \\
\botrule
\end{tabular}}
\label{table_minimal}
\end{table}

\clearpage
\begin{table}[tbp]
\tbl{Values of the mean-squared helicity based on the smoothed velocity and vorticity, $\ov{\tl{h}^{2}}$, for the DNS and for the NS-$\alpha$ model.}
{\begin{tabular}{@{}lcc}\toprule
model & $\ov{\tl{h}^{2}}$ \\
\colrule
DNS (minimal channel, mesh IV) & 0.31 \\
NS-$\alpha$  (minimal channel, mesh IV) & 0.37 \\
NS-$\alpha$  (damping, full channel) & 0.29 \\
no model (full channel) & 0.21 \\
\botrule
\end{tabular}}
\label{heltable}
\end{table}

\clearpage
\center

\clearpage

\clearpage
\center

\clearpage

\clearpage

\clearpage

\clearpage

\clearpage
\bibliographystyle{tJOT}
\bibliography{references,refs_books}

\end{document}